\documentclass[aps,prb,twocolumn,superscriptaddress,longbibliography]{revtex4-1}
\usepackage[colorlinks=true,citecolor=blue,linkcolor=blue,breaklinks=true]{hyperref}

\usepackage{color,graphicx}
\usepackage{bm}
\usepackage{amsmath}
\usepackage{amssymb}
\usepackage{mathrsfs}
\usepackage{times}
\allowdisplaybreaks

\begin{document}
	
\newcommand {\beq} {\begin{equation}}
\newcommand {\eeq} {\end{equation}}
\newcommand {\bqa} {\begin{eqnarray}}
\newcommand {\eqa} {\end{eqnarray}}
\newcommand {\ca} {\ensuremath{c^\dagger}}
\newcommand {\ba} {\ensuremath{b^\dagger}}
\newcommand {\Ma} {\ensuremath{M^\dagger}}
\newcommand {\psia} {\ensuremath{\psi^\dagger}}
\newcommand {\fbar} {\ensuremath{\bar{f}}}
\newcommand {\psita} {\ensuremath{\tilde{\psi}^\dagger}}
\newcommand{\lp} {\ensuremath{{\lambda '}}}
\newcommand{\A} {\ensuremath{{\bf A}}}
\newcommand{\QQ} {\ensuremath{{\bf Q}}}
\newcommand{\kk} {\ensuremath{{\bf k}}}
\newcommand{\qq} {\ensuremath{{\bf q}}}
\newcommand{\kp} {\ensuremath{{\bf k'}}}
\newcommand{\rr} {\ensuremath{{\bf r}}}
\newcommand{\rp} {\ensuremath{{\bf r'}}}
\newcommand {\ep} {\ensuremath{\epsilon}}
\newcommand{\nbr} {\ensuremath{\langle r r' \rangle}}
\newcommand {\no} {\nonumber}
\newcommand{\up} {\ensuremath{\uparrow}}
\newcommand{\dn} {\ensuremath{\downarrow}}
\newcommand{\rcol} {\textcolor{red}}
\newcommand{\bcol} {\textcolor{blue}}
\newcommand{\lt} {\left}
\newcommand{\rt} {\right}
\newcommand{\comment}[1]{}
\newcommand{\dt}{\phantom{\tiny 1}}


\title{Subgap two-particle spectral weight in disordered $s$-wave superconductors: Insights from mode coupling approach}
\author{Prathyush P. Poduval}\email{prathyushp@iisc.ac.in}
\affiliation{ Indian Institute of Science, Bangalore 560012, India}
\author{Abhisek Samanta}\email{abhiseks@campus.technion.ac.il}
\affiliation{ Physics Department, Technion, Haifa 32000, Israel}
\author{Prashant Gupta}\affiliation{Department of Physics, University of Illinois, Chicago 60607, USA}
\affiliation{UM-DAE Centre for Excellence in Basic Sciences (CEBS), Mumbai 400098, India}
\author{Nandini Trivedi}
\affiliation{ Department of Physics, The Ohio State University, Columbus, Ohio, USA 43201}
\author{Rajdeep Sensarma}\email{sensarma@theory.tifr.res.in}
\affiliation{ Department of Theoretical Physics, Tata Institute of Fundamental Research, Mumbai 400005, India}

\date{\today }

\begin{abstract}
We study the two-particle spectral functions and collective modes of weakly disordered superconductors using a disordered attractive Hubbard model on square lattice. We show that the disorder induced scattering between collective modes leads to a finite subgap spectral weight in the long wavelength limit. In general, the spectral weight is distributed between the phase and the Higgs channels, but as we move towards half-filling the Higgs contribution dominates. The inclusion of the density fluctuations lowers the frequency at which this mode occurs, and results in the phase channel gaining a larger contribution to this subgap mode. Near half-filling, the proximity of the system to the charge density wave (CDW) instability leads to strong fluctuations of the effective disorder at the commensurate wave-vector ($[\pi,\pi]$). We develop an analytical mode coupling approach where the pure Goldstone mode in the long wavelength limit couples to the collective mode at $[\pi,\pi]$. This provides insight into the location and distribution of the two-particle spectral weights between the Higgs and the phase channels. 
\end{abstract}


\pacs{PACS}

\maketitle

\section{Introduction}
The superconductor-insulator transition in two dimensional films
as a function of disorder strength is one of the most studied
quantum phase transitions in nature~\cite{Shahar,Sacepe,Avishai}. There is strong experimental~\cite{Sacepe,Avishai,Pratap} and theoretical evidence~\cite{Nandini1, Nandini2,Nandini3} that the transition is driven by the disordering of the phase of the Cooper pairs, rather than by weakening of amplitude of their formation. Several interesting phenomena, including the existence of single particle gap~\cite{Pratap,Sacepe}, high frequency inductive electric response~\cite{Pratap} and high magnetoresistance~\cite{Sambandamurthy} across the transition support the idea that 
the Cooper pairs exist across the transition. Hence the focus naturally shifts to the properties of the two-particle spectral functions, which include the low energy collective fluctuations of the superconducting order parameter. The fluctuations of the phase and the amplitude of the order parameter constitute the low energy spectral weight in a clean superconductor. The long wavelength amplitude mode, related to the Higgs excitation~\cite{HiggsRevShimano,VermaPekker} of high energy physics~\cite{ATLAS}, has been observed through non-linear spectroscopy~\cite{HiggsExpt,matsunaga2013higgs,matsunaga2014light,matsunaga2017polarization,cea2016nonlinear,seibold2021third}.

A key observation from recent experiments~\cite{Frydmannature} is the availability of optical spectral weight well below the two particle continuum in a disordered superconductor. The presence of this low energy spectral weight together with a hard single-particle gap was surprising. It is well known that in a clean superconductor, the long wavelength collective mode occurs at zero frequency (Goldstone mode), and there is no spectral weight until one reaches the two-particle continuum threshold at twice the single particle gap~\cite{VermaPekker,VermaLittlewood,Randeriabroken, Diener}.  Recent theoretical work~\cite{HiggsAbhisek,Benfatto1,Benfatto2,ThermalHiggsAbhisek}, which constructs the collective modes around the disordered mean-field solutions~\cite{Xiang,Nandini1,Atkinson} of a Bogoliubov de-Gennes theory, has shown the presence of spectral weight of two-particle excitations at finite frequencies below the continuum threshold.
\begin{figure}[t!]
	\centering 
	\includegraphics[width=0.485\textwidth]{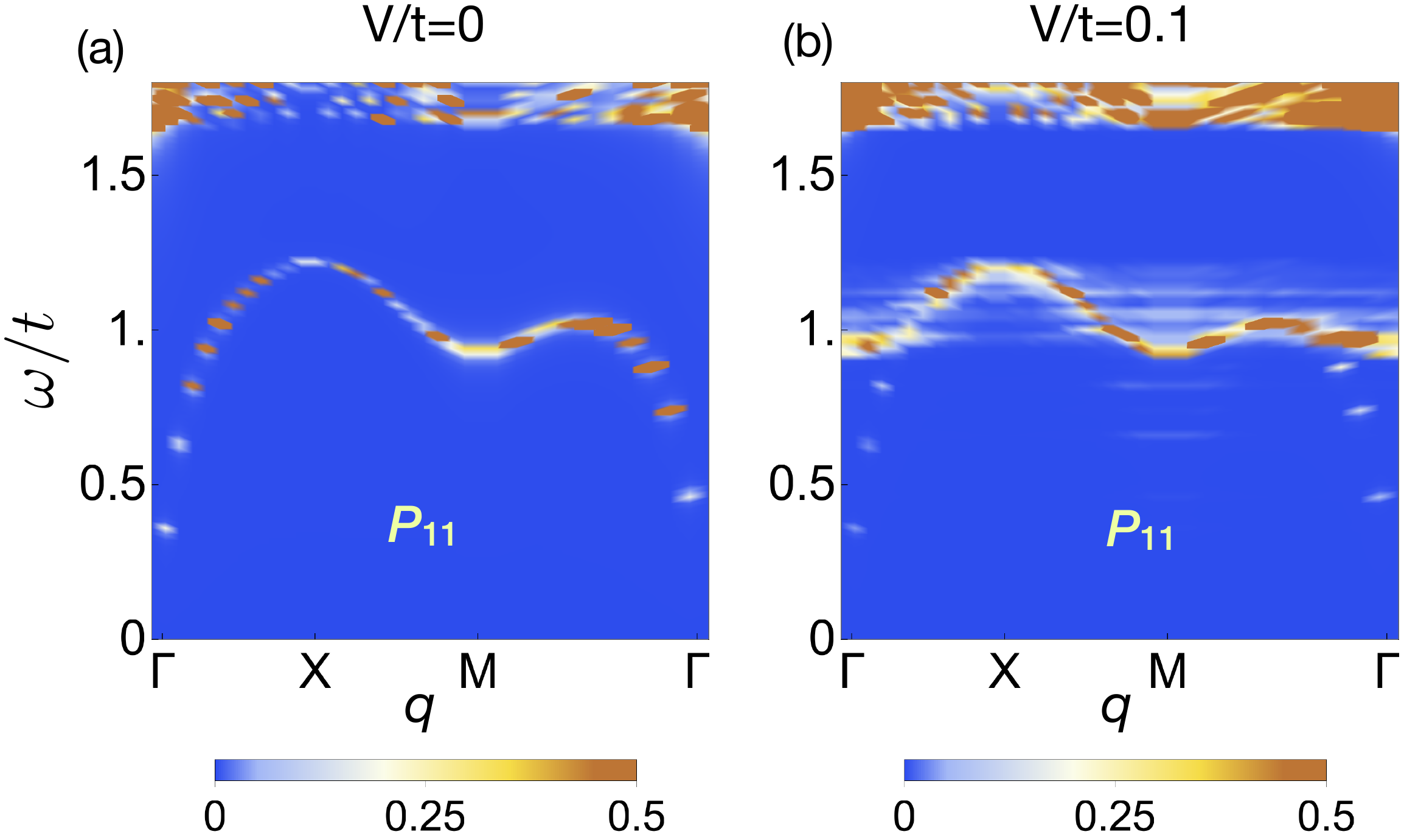}
	\caption{The disorder averaged amplitude spectral function $P_{11}(\qq,\omega)$ for a system with (a) $V/t=0$ corresponding to the clean case and (b) $V/t=0.1$ corresponding to weak disorder. The spectral function are calculated by expanding the action around the inhomogeneous  BdG saddle point. In (b), a subgap mode is clearly seen at $\omega/t\sim ~1$ around the $\Gamma$ point ($\qq=[0,0]$). This is calcualted for a $24\times 24$ lattice with $U/t=3$ and $n=0.875$ (Data taken from Ref.~\onlinecite{HiggsAbhisek}).}
	\label{fig:higgssubgap}
\end{figure}

An important feature of the theoretical results is that the subgap weight of two-particle excitations exists even at the weakest disorder, showing that this is not a feature which can only be associated with the quantum phase transition and shows a non-perturbative (in disorder) redistribution of the spectral weight in the long wavelength limit. The details of the subgap spectral weight however crucially depends on the approximations used in the theoretical calculations: In Ref.~\onlinecite{HiggsAbhisek,ThermalHiggsAbhisek}, the authors only considered the fluctuations of the pairing field and obtained a subgap feature which had a narrow spectral range and was dominated by the amplitude or Higgs component. In contrast, in Ref.~\onlinecite{Benfatto1}, the authors also considered the fluctuations of the density field, but focused only on the $\qq=[0,0]$ spectral function, where they obtained broad spectral weights dominated by the phase mode. 
\begin{figure*}[t]
	\centering 
	\includegraphics[width=0.75\textwidth]{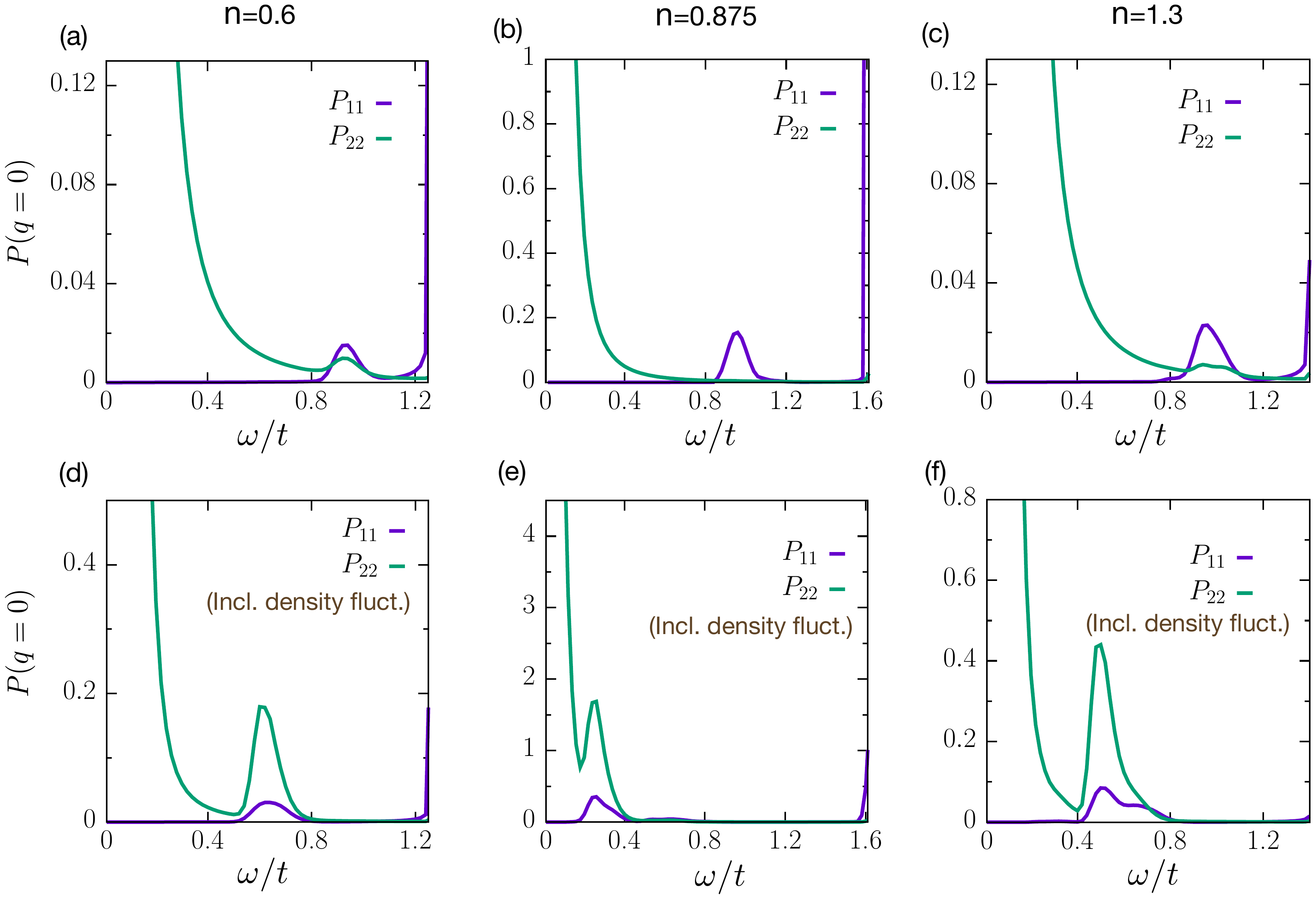}
	\caption{The disorder averaged energy distribution curves (EDC) for amplitude and phase spectral functions ($P_{11}$ and $P_{22}$) for $\qq=[0,0]$ for different densities: For (a) and (d): $n=0.6$,  (b) and (e): $n=0.875$, (c) and (f): $n=1.3$. The top row shows the spectral functions without the effects of dynamic density fluctuations~\cite{HiggsAbhisek} and the bottom row shows the spectral functions after including the effects of dynamic density fluctuations~\cite{Benfatto1}. The results from expansion around the inhomogeneous saddle point show that: (i) The phase contribution to subgap mode increases as we move away from half-filling on either side (a-c) (ii) On including density fluctuations, the location of the subgap mode shifts to a lower value of $\omega$ and the phase contribution to the subgap mode increases substantially (d-f). All the data are for a $24\times 24$ lattice with $U/t=3$ and $V/t=0.1.$ }
	\label{fig:exactdisorder}
\end{figure*}

In this paper, we understand the systematic trends of the subgap two-particle spectral weight in the weakly disordered $s$-wave superconductor using the disordered attractive Hubbard model on a square lattice as a prototype. To get analytic insights, we consider the mean-field saddle point of the translation invariant system (with no disorder) and expand the theory both in terms of static fluctuations created by the disorder and dynamic quantum fluctuations of the pairing and density fields. We thus obtain a description in terms of translation invariant collective modes being scattered by an effective disorder. Using a simple Born approximation to account for the disorder scattering, we show that we can reproduce the subgap feature seen in earlier works. We also find that the static fluctuations are peaked around the commensurate vector of $\QQ=[\pi,\pi]$, a reflection of the charge density wave instability of the square lattice Hubbard model at half-filling. Although the theoretical calculations are done away from half-filling, the proximity effect causes this peak once the translation symmetry is broken by disorder. This motivates us to consider a simplified mode coupling theory, where the collective modes at $\qq$ are coupled to those at $\qq+\QQ$. 

Using this mode coupling theory, we show that the two-particle spectral weight at finite subgap frequencies at $\qq=[0,0]$  originates from the scattering of the $\QQ=[\pi,\pi]$ mode by the effective disorder. Hence, at weak disorder, this weight appears around the energy of the $[\pi,\pi]$ mode in the clean system. We find that the particle-hole symmetry at half-filling ensures that the Goldstone mode at $\qq=[0,0]$ couples only to the amplitude component of the $[\pi,\pi]$ mode, and hence the weight shows up only in the amplitude or Higgs channel close to half-filling. As one moves away from half-filling, the phase contribution to this subgap mode increases, as seen in the numerical calculations. The inclusion of dynamic density fluctuations~\cite{Benfatto1} lowers the collective mode frequency at $[\pi,\pi]$. As a result, the subgap weight is shifted to lower frequencies, and the spectral separation between this mode and the tail of the low energy weight from the Goldstone mode is lost. Further, this lowering of the energy also implies that the subgap weight has a larger mixing of the pure phase Goldstone mode; thus, the contribution of the phase component dominates in this case.

We note that if a disordered superconductor is close to a charge density wave transition, as in NbSe$_2$~\cite{NbSe2cdwexpt}, the strong static fluctuations at the commensurate wave-vector will dominate the disorder scatterings. The presence of subgap spectral weight in the two-particle spectral function will also be a generic feature in that case. We know that unlike the work of Varma and Littlewood~\cite{VermaLittlewood,VermaPekker}, where the system has additional CDW order, here the system is close to but not in the CDW phase. Hence in a clean system, there will be no subgap weight at $\qq=[0,0]$. However, the presence of disorder, which breaks translational symmetry and allows for scattering of collective modes, leads to the formation of strong subgap spectral features in these systems.

The rest of the paper is organised as follows: In Section~\ref{sec2}, we provide a summary of results on the two-particle spectral function of a weakly disordered attractive Hubbard model on a square lattice,  obtained from numerical calculations using BdG theory and expansions around this inhomogeneous saddle point. In Section~\ref{sec3}, we expand the theory around the translation invariant saddle point in both the disorder induced static spatial fluctuations as well as the dynamic quantum fluctuations. This leads to a model of translation invariant collective modes scattered by an effective disorder. We show that a simple Born approximation can reproduce the subgap spectral weight. In Section~\ref{sec4}, we derive an approximate mode coupling theory by focusing on the fact that the nearby CDW instability leads to a peak in the static correlators at the corresponding commensurate wave-vector (here $[\pi,\pi]$). We then use this mode coupling theory to understand the systematic trends in the numerical calculations around the inhomogeneous mean-field solutions. We finally conclude by summarizing in Section~\ref{sec5}.

\section{Collective Modes in Disordered Superconductors: Results from Fermionic Theory}
\label{sec2}

\begin{figure}
\includegraphics[width=0.485\textwidth]{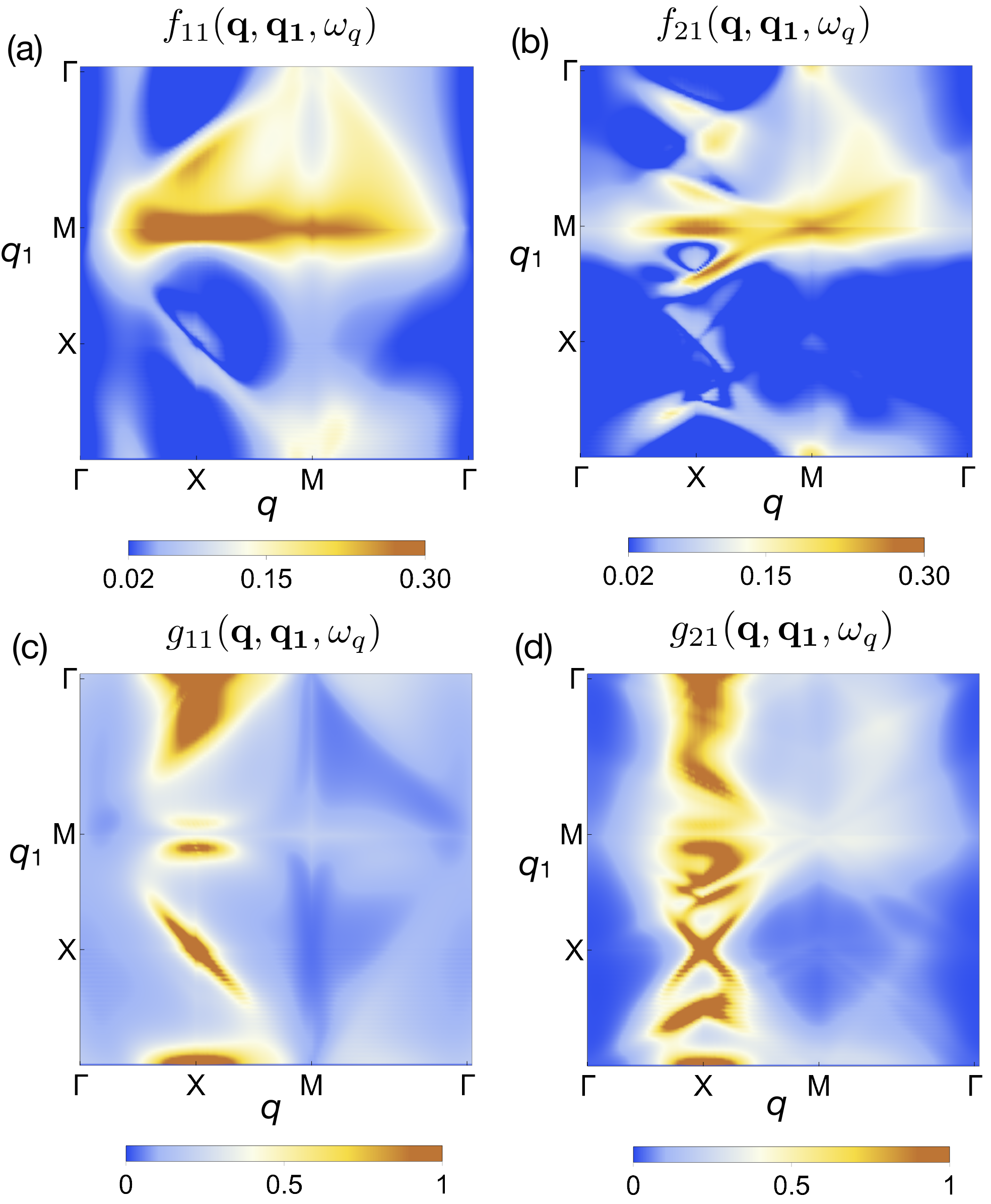}
\caption{The static fluctuations of the Hartree fields $\delta v_{\qq_1}$ scatters the collective modes at momentum $\qq$ to modes at $\qq+\qq_1$ with a coupling $f_{ij}(\qq,\qq_1,\omega_\qq)$, where $\omega_\qq$ is the dispersion of the collective mode (See Eq.~\ref{eq:cartdis} and \ref{eq:fg}). The color-plots of $f_{11}$ and $f_{21}$ as a function of $\qq$ and $\qq_1$ along the principal axes are shown in (a) and (b) respectively. The corresponding couplings for static fluctuations of pairing fields $\delta\Delta_{\qq_1}$, $g_{11}$ and $g_{21}$ are shown in (c) and (d) respectively. $f_{11}$ and $f_{21}$ are relatively independent of $\qq$ and are peaked around $\qq_1=M$. $g_{11}$ and $g_{21}$ on the other hand is independent of $\qq_1$ but peaked around $\qq=X$.}
\label{fig:fgplot}
\end{figure}
 In this section we will review the results on collective modes of disordered $s$-wave superconductors obtained from a theory of fermions with attractive interactions in the presence of a random disorder potential. This will give us the key phenomenology which we want to explain; at the same time it will help in setting up the basic theoretical framework that we will use in the rest of the paper. Some of these results have been previously reported in Ref.~\onlinecite{HiggsAbhisek},~\onlinecite{Benfatto1},~\onlinecite{Benfatto2}, and Ref.~\onlinecite{ThermalHiggsAbhisek}, while some of the systematics of the various trends are being reported here for the first time.

We work with the attractive Hubbard model on a square lattice, with random potential disorder at zero temperature, given by the Hamiltonian
\begin{equation}
\label{eq:Ham}
H=-t\sum_{\nbr\sigma}(\ca_{r\sigma}c_{r' \sigma} + h.c )-U\sum_rn_{r\up}n_{r\dn}+\sum_r (v_r-\mu)n_r
\end{equation}
where $c^{\dagger}_{r\sigma}(c_{r\sigma})$ is the creation (annihilation) operator for an electron with spin $\sigma$ on site $r$, and $\mu$ is the chemical potential. Here $t$ is the nearest neighbour hopping parameter, and $U$ is the local attractive interaction between the electrons. $v_r$ is an independent random variable for each site that is uniformly sampled from $[-V/2,V/2]$; thus $V$ characterises the scale of the disorder.

The first step is to construct a mean-field theory in terms of the pairing field $\Delta_0(r) =U \langle c^\dagger_{r\up}c^\dagger_{r\dn}\rangle$ and the Hartree shift $\xi_0(r)=U\langle c^\dagger_{r\sigma} c_{r\sigma}\rangle$. This leads to the mean-field Bogoliubov de-Gennes Hamiltonian
\beq
\left( \begin{array}{cc}%
         H_0(rr')& \Delta_0(r) \delta_{rr'}\\
         \Delta_0(r) \delta_{rr'} & -H_0(rr')
                                    \end{array}\right) \left( \begin{array}{c}%
           u_m(r') \\
          v_m(r') \end{array}\right) =E_m \left( \begin{array}{c}%
           u_m(r) \\
          v_m(r) \end{array}\right)
\eeq
where $H_0(rr')=-t_{rr'}-[\mu-v^{eff}(r)]\delta_{rr'}$ if $r$ and $r'$ are either nearest neighbours or the same site, and $0$ otherwise. Here $t_{rr'}=t$ if $r$ and $r'$ are nearest neighbours and $0$ otherwise. The microscopic disorder potential $v_r$ is renormalized by the Hartree shift to the effective disorder potential $v^{eff}(r)= v_r-\xi_0(r)$. 
We note that while $v_r$ is an independent random variable for each site, $\Delta_0(r)$ and $v^{eff}(r)$ for different sites have finite correlations between them. The self-consistent mean field equations at $0$ temperature are then given by
\begin{eqnarray}
  \no \Delta_0(r) &=&U\!\!\!\sum_{m:E_m>0} u_m(r)v_m(r)\\
  \xi_0(r) &=& U\!\!\!\sum_{m:E_m>0} v^2_m(r)
  \end{eqnarray}
Additionally, we fix the average density of each disorder configuration to $n$ by solving the number equation $n=\frac{2}{N_s}\sum_r \sum_{m:E_m>0} v_m^2(r)$. It is well known from earlier works~\cite{Nandini1,Yenleeloh,Debmalya}, that the distributions of $\Delta_0(r)$ and $v^{eff}(r)$ change from a narrow distribution around the mean value at low disorders to bimodal distributions indicating the formation  of superconducting and non-superconducting patches in the system at strong disorder. 

The collective modes in disordered superconductors arise from the spatio-temporal \textit{fluctuations} of the pairing field about the inhomogeneous mean field solution. In an imaginary time ($\tau$) formalism, this is achieved by  considering 
$\Delta(r,\tau)=\left(\Delta_0(r)+\eta(r,\tau)\right)e^{i\theta(r,\tau)}$,
where $\eta(r,\tau)$ is the amplitude  and $\theta(r,\tau)$ is the phase fluctuation. Expanding the action to second order in these fluctuation fields, one obtains a non-interacting theory (quadratic action) of the amplitude and phase fluctuations,
\beq
S=\!\!\! \sum_{rr',\omega_n}\!\! (\eta(r,i\omega_n), \theta(r,i\omega_n))\hat{D}^{-1}(r,r'\!,i\omega_n)\!\!\left(\! \begin{array}{c}%
           \eta(r',-i\omega_n) \\
                                                                \theta(r',-i\omega_n) \end{array}\!\right)
                                               \label{quadaction}             \eeq
where the details of the $2\times 2$ matrix inverse propagator $\hat{D}^{-1}$ is given in Ref.~\onlinecite{HiggsAbhisek}. The experimentally measurable amplitude spectral function is given by the analytic continuation to real frequencies, $P_{11}(r,r',\omega) = -\frac{1}{\pi} \textrm{Im} D_{11}(r,r',i\omega_n \rightarrow \omega+i0^+)$, while the corresponding phase spectral function is given by $P_{22}(r,r',\omega) = -\frac{1}{\pi} \Delta_0(r)\Delta_0(r') \textrm{Im} D_{22}(r,r',i\omega_n \rightarrow \omega+i0^+)$. In a disordered system, the spectral functions, calculated for a particular disorder configuration, does not have  translational invariance. However, translation invariance is restored on disorder averaging, so that the disorder averaged spectral function can be Fourier transformed in spatial co-ordinates. This disorder averaged spectral functions, $P_{11}(\qq,\omega)$ and $P_{22}(\qq,\omega)$, have been studied in detail previously as a function of disorder at zero temperature~\cite{HiggsAbhisek,Benfatto1} and at finite temperatures~\cite{ThermalHiggsAbhisek}. In this paper, we will only present results for zero temperature.
\begin{figure*}
\centering 
	\includegraphics[width=0.99\textwidth]{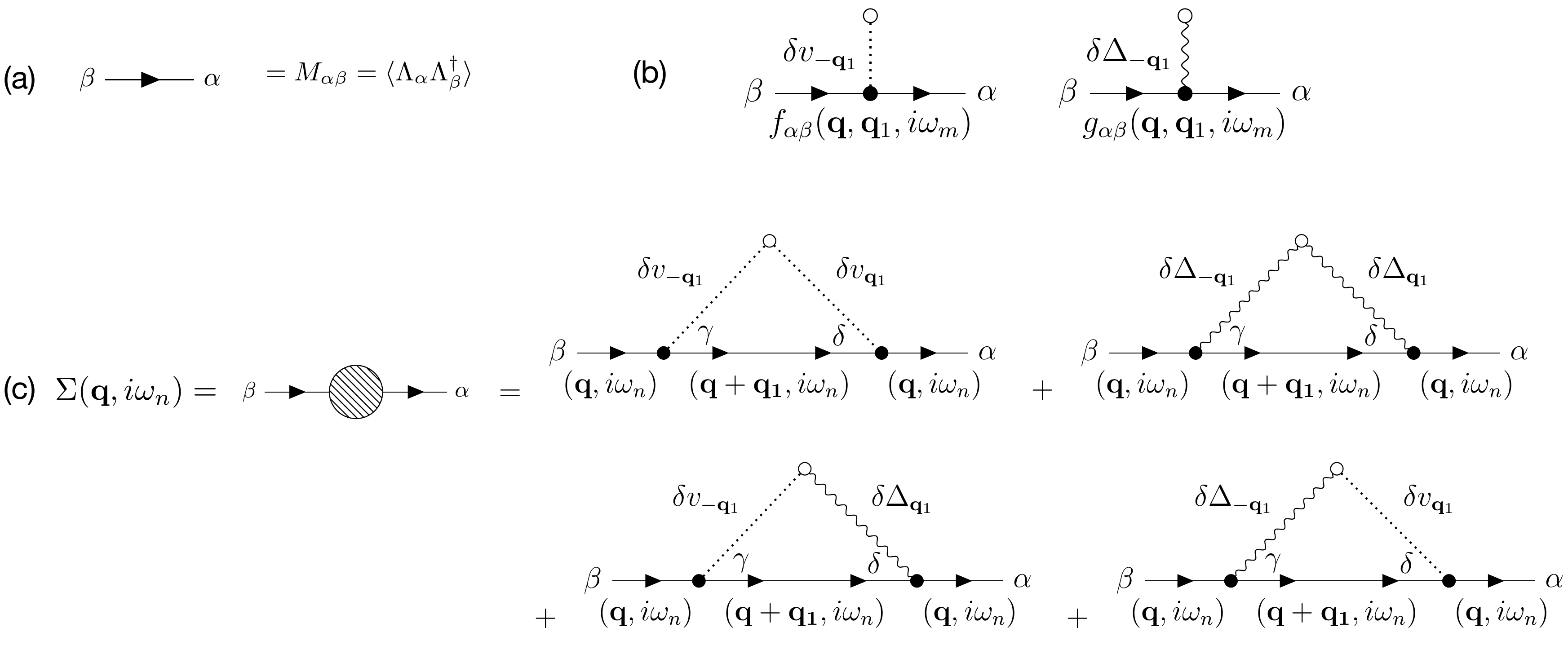}
\caption{The Feynman diagrams for the self-energy of the collective modes due to scattering by the static fluctuations. (a) The line denoting the propagator of quantum fluctuations (b) The Feynman vertices that couple the static fluctuations with the dynamic fluctuations (c) The self-energy diagrams in the leading order Born approximation. The self-energy terms are proportional to $\langle|\delta\Delta(\qq)|^2\rangle,\langle|\delta v(\qq)|^2\rangle$ and $\langle\delta v(\qq)\delta\Delta(-\qq)\rangle$.}
\label{fig:feynmandiag}
\end{figure*}

An intriguing result from Ref.~\onlinecite{HiggsAbhisek} is the dramatic change in the low energy Higgs (amplitude) spectral function at weak disorder. In the clean case, it is well known that the linearly dispersing collective mode has a pure phase character as $ \qq\rightarrow [0,0]$, i.e., the amplitude component goes to $0$. At $\qq=[0,0]$, the Higgs spectral weight resides at the two-particle continuum threshold and the mode is damped out. In contrast, even at very weak disorder, there are sharply defined excitation modes (peaks in spectral function) in the Higgs channel below the two-particle continuum in the long-wavelength limit  ($\qq=[0,0]$). In Fig~\ref{fig:higgssubgap}, we show the color-plot of the amplitude spectral function $P_{11}(\qq,\omega)$ in the $\qq-\omega$ plane (with $\qq$ along the principal axes in the Brillouin zone), calculated from such a Gaussian expansion around the mean-field theory at $U/t=3$ and $n=0.875$. Fig.~\ref{fig:higgssubgap}(a) shows the clean case ($V/t=0$) where the Higgs weight at $\qq=[0,0]$ ($\Gamma$ point) starts from the two-particle continuum threshold. Fig.~\ref{fig:higgssubgap}(b)  shows the spectral function at a weak disorder of $V/t=0.1$. In this case, it is clear that at $\qq=[0,0]$, the amplitude spectral function has finite weight at an energy $\omega_0$, which is below the two-particle gap $2\Delta_0$. Surprisingly, the weight is observed only over a narrow band of frequencies close to the collective mode frequency at $\QQ=[\pi,\pi]$, which suggests that the weight signifies an actual quasiparticle excitation. This is unlikely to be caused just by the incoherent scatterings from the disorder potential, and suggests the existence of a more fundamental mechanism. In addition, we also see a small but finite low energy weight near the $M$ point, which was absent in the clean case. We note that we will present data for $U/t=3$ in this paper unless otherwise mentioned.

In order to understand the systematic changes in the long-wavelength spectral functions, we plot the energy distribution curves (EDCs) at $\qq=[0,0]$ for the amplitude and phase channel in Fig.~\ref{fig:exactdisorder}. Fig.~\ref{fig:exactdisorder} (a)-(c) shows the results at a weak disorder of $V=0.1t$ for three different densities, $n=0.6$, $n=0.875$ and $n=1.3$ respectively. In all the cases, we see that the subgap spectral weight around $\omega_0$ is spectrally well separated from the low energy phase contribution of the collective mode. The spectral weight around $\omega_0$ is exclusively in the Higgs channel at $n=0.875$, whereas the weight is more evenly distributed between the amplitude and phase channels as we move away from half-filling ($n =0.6$ and $n=1.3$). This seems to suggest that some approximate symmetry suppresses the phase contribution in this case as one approaches half-filling.

Finally, in Ref.~\onlinecite{Benfatto1}, the fluctuations of the Hartree (local density) field $\xi(r,\tau)$ were considered along with those of the pairing field. Expanding up to quadratic order, one gets an action similar to Eq.~\ref{quadaction}, now with a three component field $(\eta,\theta,\xi)$ and a $3\times 3$ matrix propagator. The density fluctuations were then integrated out to obtain the effective collective modes for the $\eta$ and $\theta$ fluctuations in the system. It was found that the subgap weight is shifted to much lower energy and overlaps with the spectral weight from the Goldstone (phase) mode. The systematic changes in the EDC on adding density fluctuations are shown in Fig.~\ref{fig:exactdisorder}(d)-(f) for densities $n=0.6,~0.875$,  and $1.3$ respectively. Although the subgap mode does survive, its location and composition change drastically. Initially, the subgap mode was part of a flat band near the top of the collective mode spectrum. But including the effects of density fluctuations causes the location of the mode shift towards $\omega=0$. Moreover, the composition of the subgap mode is dominated by the phase contributions. 

To summarize, the following effects are seen in these calculations: (1) The presence of a weak disorder seems to give rise to an excitation at a finite $\omega$ below the two-particle continuum. (2) This mode is observed to be purely in the amplitude channel for $n\sim0.875$, with the phase contribution increasing both as we move away from half-filling and when we include density fluctuations. (3) The location of this mode in $\omega$ space was initially part of a flat mode, however it gets drastically pulled down when density fluctuations are included. While these trends are clear, it is hard to obtain additional insights from these calculations, since they can only be accessed by large scale numerics. In the next section, we will formulate this problem in terms of the collective modes of the translation invariant system interacting with an effective disorder to get additional insight into these trends.

\section{Collective Bosons and their Effective Disorder}
\label{sec3}

In the fermionic theory, the disorder is modelled by a random potential felt by the electrons at every site.
The disorder strength (the width of the probability distribution of the random potential) can be related to
experimentally observable quantities like sheet resistance~\cite{Anushree,Baturina,Steiner,Pratap}, and one can make a detailed comparison with
realistic systems. However, there is a price to pay for this exact treatment of the microscopic disorder.
The eigenstates of the mean-field theory can only be determined numerically and varies from one disorder
configuration to another. Hence calculations of collective modes are numerically expensive and are limited
to small system sizes. Further, it is hard to get any insight behind the observed phenomena.

To circumvent these difficulties and obtain analytic insight into the subgap spectral weight, we obtain a 
description where the translation invariant
collective modes are scattered by an effective disorder. To achieve this, we note that within the
mean-field BdG theory, the disorder gives rise to a local pairing $\Delta_0(r)$ and an effective
local potential (microscopic disorder potential, renormalized by Hartree shifts) $v^{eff}(r)$. We
can treat these as new random variables, which determine the properties of the mean-field
solutions as well as the spectral properties of the collective modes. We break them up into
an average ($\Delta_0$ and $v_0$) and a static spatial fluctuation
($\delta \Delta_0(r)= \Delta_0(r)- \Delta_0$ and $\delta v(r)=v^{eff}(r)-v_0$). Note that, by
construction $\delta \Delta_0(r)$ and $\delta v(r)$ are correlated random variables with zero mean.
We first consider a translation invariant saddle point with $\Delta_0$ and $v_0$, where the
fermion Green's function in the Nambu basis in momentum space is given by
\begin{widetext}
\beq
G(\kk,i\omega_n)=\left(\begin{array}{cc }%
                         \frac{u_\kk^2}{i\omega_n-E_\kk}+ \frac{v_\kk^2}{i\omega_n+E_\kk} & u_\kk v_\kk\left[\frac{1}{i\omega_n-E_\kk}-\frac{1}{i\omega_n+E_\kk}\right]\\
                         u_\kk v_\kk\left[\frac{1}{i\omega_n-E_\kk}-\frac{1}{i\omega_n+E_\kk}\right]&\frac{v_\kk^2}{i\omega_n-E_\kk}+ \frac{u_\kk^2}{i\omega_n+E_\kk} \end{array}\right)
\eeq
\end{widetext}
where $\omega_n = (2n+1)\pi T$ with integer $n$ is the fermionic Matsubara frequency at temperature $T$. Here $E_\kk=\sqrt{(\epsilon_\kk-\mu+v_0)^2+\Delta_0^2}$ is the Bogoliubov quasiparticle dispersion,
$\epsilon_\kk=-2t (\cos k_x +\cos k_y)$ is the bare band dispersion, and
the BCS coherence factors are given by $u_\kk^2=1-v_\kk^2=(1/2)(1+(\epsilon_\kk-\mu+v_0)/E_\kk)$,
and $u_\kk v_\kk= \Delta_0/2E_\kk$.

Our model of translation invariant bosons coupled to effective disorder is obtained by considering
$\Delta(r,\tau) = \Delta_0+\delta \Delta(r) +\lambda(r,\tau)$ and $v(r,\tau)=v_0+\delta v(r) + \xi(r,\tau)$, and expanding the action both in the dynamic quantum fluctuations ($\lambda(r,\tau)$, $\xi(r,\tau)$) and the disorder induced static fluctuations $\delta \Delta(r)$ and $\delta v(r)$. This leads to the fluctuation action
\begin{widetext}
\beq
\label{eq:cartdis}
S_{fl}=\frac{1}{2}\sum_{\qq,\omega_m} \Lambda^\dagger(\qq,i\omega_m) \hat{M}^{-1}(\qq,i\omega_m)\Lambda(\qq,i\omega_m) +\sum_{\qq,\qq_1,\omega_m} \Lambda^\dagger(\qq+\qq_1,i\omega_m) \hat{\tilde{F}}(\qq,\qq_1,i\omega_m)\Lambda(\qq,i\omega_m) 
\eeq
where $\omega_m=2m\pi T$, with integer $m$, is the bosonic Matsubara frequency, and the three-component field $\Lambda^\dagger(\qq,i\omega_m)=[\lambda^\ast(\qq,i\omega_m),\lambda(-\qq,-i\omega_m),\xi(\qq,i\omega_m)]$. We note that when we will analyse the theory without density fluctuations, we will set $\xi=0$ and work with a two-component field.

Here, $\hat{M}^{-1}$ is the inverse propagator for the translation invariant collective modes,
\bqa
 M^{-1}_{ij}(\qq,i\omega_m)&=& \frac{\delta_{ij}(1+\delta_{i3})}{U} + \sum_{\kk,\omega_n} \text{Tr }^NG(\kk+\qq,i\omega_n+i\omega_m)\sigma^{\alpha_i} G(\kk,i\omega_n)\sigma^{\beta_j}
\eqa
where the trace is over Nambu indices, $\alpha_i=+,-,3$ for $i=1,2,3$ respectively and $\beta_j=-,+,3$ for $j=1,2,3$ resepctively. Here $\sigma^{i=1,2,3}$ denote the Pauli matrices and $\sigma^\pm = \sigma^1\pm i\sigma^2$. The detailed evaluation of $M^{-1}$ is given in Appendix~\ref{app:cartesian}  (also see Ref.~\onlinecite{Randeriabroken,Diener,Benfatto1} for earlier
derivations of the propagator). The low energy poles of $\hat{M}(\qq,\omega+i0^+)$ determine the collective mode
frequencies of the translation invariant system, which disperse linearly at low momenta.
The second term in the action scatters a fluctuation at momentum $\qq$ to a fluctuation at $\qq+\qq_1$
(with the same frequency), and is linearly dependent on the (Fourier transformed) static fluctuations
$\delta \Delta_{\qq_1}$ and $\delta v_{\qq_1}$. The scattering matrix $\hat{\tilde{F}}$ can be written as
  \bqa
\label{eq:fg}
\hat{\tilde{F}}_{ij}(\qq,\qq_1,i\omega_m)&=&f_{ij}(\qq,\qq_1,i\omega_m)\delta \Delta_{-\qq_1}+ g_{ij}(\qq,\qq_1,i\omega_m)\delta v_{-\qq_1}
\eqa
  where $i,j$ run between $1$and $3$. The coupling functions are given by
 \bqa
\no f_{ij}(\qq,\qq_1,i\omega_m)&=& \sum_{\kk,\omega_n} \text{Tr }^NG(\kk+\qq+\qq_1,i\omega_n+i\omega_m)\sigma^{\alpha_i} G(\kk,i\omega_n)\sigma^{1}G(\kk+\qq_1,i\omega_n)\sigma^{\beta_j}\\
 g_{ij}(\qq,\qq_1,i\omega_m)&=&\sum_{\kk,\omega_n} \text{Tr }^NG(\kk+\qq+\qq_1,i\omega_n+i\omega_m)\sigma^{\alpha_i} G(\kk,i\omega_n)\sigma^{3}G(\kk+\qq_1,i\omega_n)\sigma^{\beta_j}
\eqa
\end{widetext}

Other than the matrix structure, these terms have the general interpretation of an effective disorder
scattering the collective modes at low energy.
The explicit evaluation of the coupling functions $f_{\alpha\beta}$ and $g_{\alpha\beta}$ is shown in Appendix~\ref{app:cartesian}. For real frequencies below the two-particle continuum, these coupling functions are real.
Since we are interested in the scattering of the collective modes due to disorder, the coupling of interest
to us is obtained by considering the analytic continuation $i\omega_m \rightarrow \omega_\qq+i0^+$,
where $\omega_\qq$ is the collective mode frequency at $\qq$. These couplings are the amplitude for scattering a
collective mode at $\qq$ by a momentum $\qq_1$. In Fig.~\ref{fig:fgplot}(a), we plot $f_{11}(\qq,\qq_1,\omega_\qq) $
as a color-plot in the $\qq-\qq_1$ plane (with momenta taken along principal axes of the Brillouin zone).
We find that the coupling is peaked around the momentum transfer of $\qq_1 =\QQ=[\pi,\pi]$,
  while it is reasonably independent of the starting wave-vector $\qq$. In Fig.~\ref{fig:fgplot}(b)-(d), we
plot the coupling functions $f_{21}$, $g_{11}$ and $g_{21}$ respectively.  We find that $f_{11}$ and $f_{21}$ have
 similar dependence on momenta. On the contrary,  $g_{11}$ and $g_{21}$  are very sensitive to $\qq$, peaking around 
 $\qq=X=[\pi,0]$, and are relatively independent of $\qq_1$, the momentum transferred in the scattering.
The scattering of the quantum fluctuations around the translation invariant saddle point by the static
disorder can be represented in terms of Feynman diagrams, with fluctuation propagators and 
vertices coupling the dynamic fluctuations to the static effective disorder, as shown in  Fig.~\ref{fig:feynmandiag}.
We note that this action is derived to leading order in the static fluctuations, and hence would fail
to account for higher order scatterings at strong disorder. However, we are only interested in the
properties of the collective modes at weak disorder; so this provides a sufficient starting point to
understand the phenomenology described in the earlier section.

\begin{figure}
\includegraphics[width=0.485\textwidth]{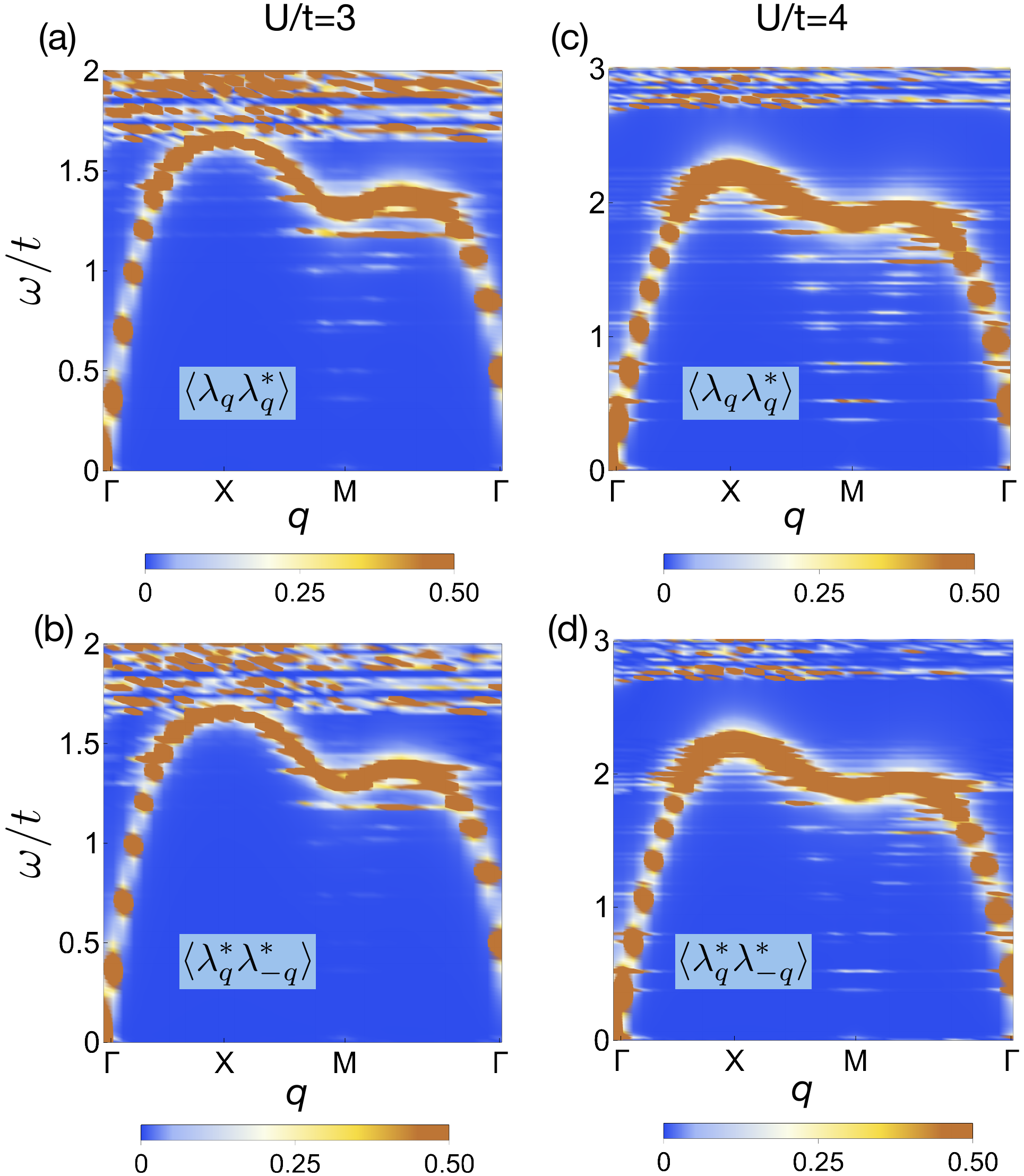}
\caption{The fluctuation spectral functions calculated within the Born approximation with effective disorder obtained from mean-field BdG solutions for (a) and (b) $U/t=3$ and (c) and (d) $U/t=4$. Here we use the cartesian form of the fluctuations. (a) and (c) give the diagonal part of the spectral function and (b) and (d) give the off-diagonal parts. There is a small pileup of weight at $\qq=[0,0]$ signifying the formation of a subgap mode.}
\label{fig:borncartesianfull}
\end{figure}

After constructing the effective action for the collective bosons and their scattering due to disorder,
we need an approximate way of incorporating the effects of these scatterings into the spectral
function of the fluctuations. This is done by dressing the inverse propagator $M^{-1}$ by the
self-energies due to disorder scattering $M^{-1} \rightarrow M^{-1}-\Sigma$, and constructing
the imaginary part of the propagator obtained from this dressed Green's function for the
fluctuations. For this, we use the simplest Born approximation scheme (the self-energy diagrams are shown in Fig.~\ref{fig:feynmandiag}), where the self-energy matrix can be written as

 \beq
 \hat{\Sigma}(\qq,i\omega_m)= \sum_{\qq_1} \hat{\tilde{F}}_{\qq+\qq_1,-\qq_1,i\omega_m} \hat{M}_{\qq+\qq_1,i\omega_m} \hat{\tilde{F}}_{\qq,\qq_1,i\omega_m}
 \eeq

 We note that there is a self-energy term linear in $\hat{\tilde{F}}$, which vanishes on averaging over disorder,
 leaving this as the leading order contribution. The Born approximation is valid at weak
 disorder and will fail to capture non-perturbative effects due to disorder scattering.

\begin{figure*}
	\includegraphics[width=0.85\textwidth]{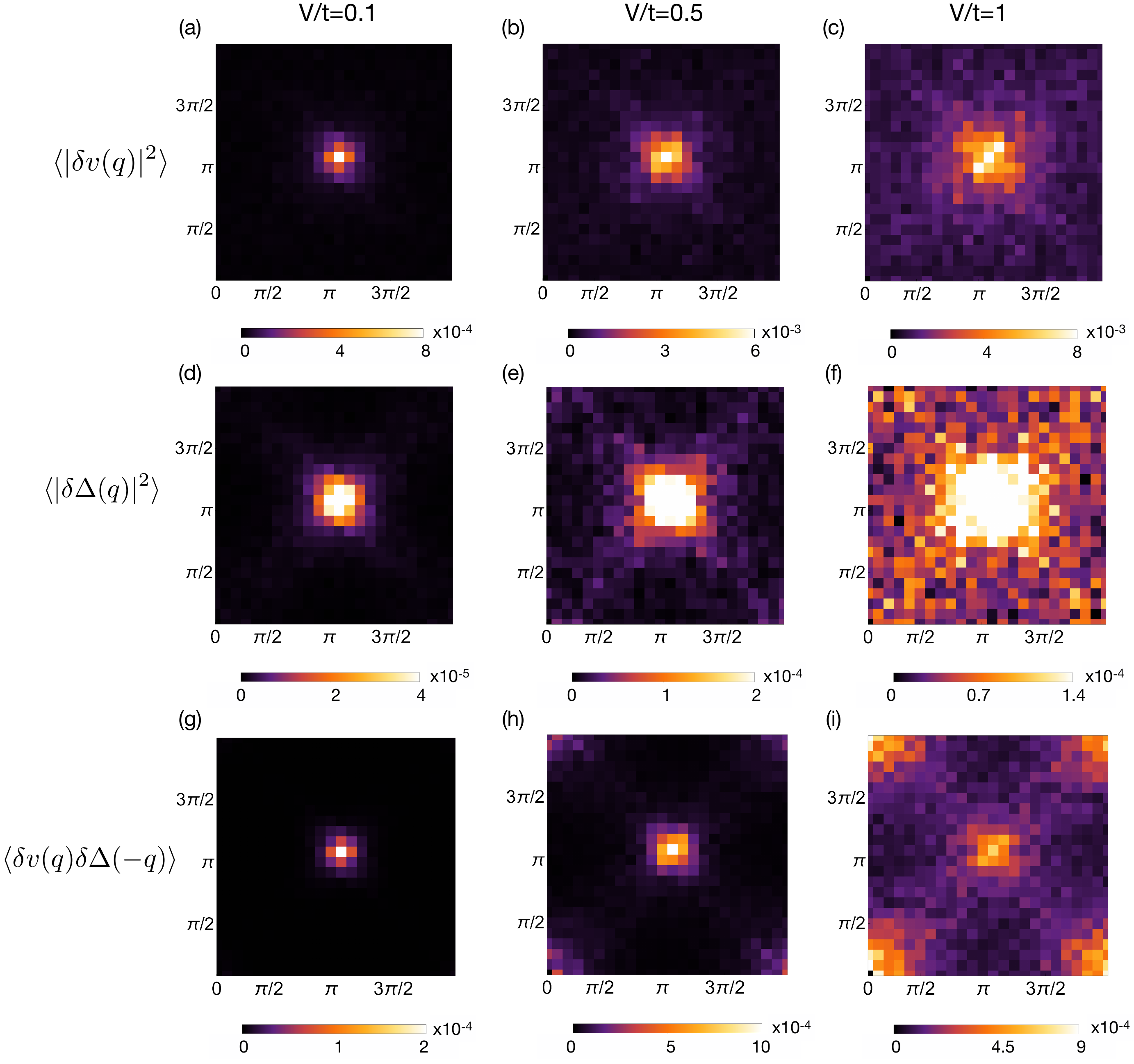}
	\caption{The disorder averaged static fluctuations of (a)-(c): The Hartree field $\langle|\delta v(\qq)|^2\rangle$, with $V/t=0.1$ for (a), $0.5$ for (b) and $1.0$ for (c). (d)-(f): The pairing field $\langle|\delta\Delta(\qq)|^2\rangle$, with $V/t=0.1$ for (d), $0.5$ for (e) and $1.0$ for (f). (g)-(i): The cross-correlator $|\langle\delta v(\qq)\delta\Delta(-\qq)\rangle|$, with $V/t=0.1$ for (g), $0.5$ for (h) and $1.0$ for (i) over the full $2D$ Brillouin zone. This is calculated from BdG solutions on a $24\times 24$ lattice in a system with $U/t=3$, $n=0.875$ and averaged over $15$ disorder realisations. The Hartree correlator is very strong at $(\pi,\pi)$, and then spreads out around this point as the disorder strength is increased. The superconducting correlators follow a similar pattern, however the corresponding magnitudes are about a factor of $10$ smaller than those of Hartree correlators.}
	\label{fig:correlator1}
\end{figure*}
To see the effectiveness of our approximations in capturing the phenomena described in the previous section,
we consider the static correlators
\begin{align}
\no\langle |\delta \Delta(\qq)|^2\rangle &= \frac{1}{N_s}\sum_{r,r'}e^{i\qq(r-r')} \langle \delta \Delta(r)\delta \Delta(r')\rangle\\
\no\langle |\delta v(\qq)|^2\rangle &= \frac{1}{N_s}\sum_{r,r'}e^{i\qq(r-r')} \langle \delta v(r)\delta v(r')\rangle\\
\langle\delta v(\qq)\delta\Delta(-\qq)\rangle &= \frac{1}{N_s}\sum_{r,r'}e^{i\qq(r-r')} \langle \delta v(r)\delta \Delta(r')\rangle
\end{align}
where $N_s$ is the number of sites in the system and the correlators are calculated in the inhomogeneous BdG solutions of the disordered system. Here, the averaging is over disorder realizations. These correlators then give the disorder averaged self-energy, which are used to construct the spectral functions of the fluctuations. 

We first consider a superconductor at a density $n=0.875$ at a weak disorder $V/t=0.1$ on a $24\times 24$ lattice. We suppress the quantum density fluctuations and work in the two-component formalism.  The spectral function for the $\lambda$ fluctuations in this case are plotted in Fig.~\ref{fig:borncartesianfull}. For a system with $U/t=3$, Fig.~\ref{fig:borncartesianfull}(a) shows the spectral function corresponding to $\langle \lambda_\qq \lambda_{\qq}^*\rangle$, while Fig.~\ref{fig:borncartesianfull}(b) shows spectral weights in $\langle \lambda_\qq^*\lambda_{-\qq}^*\rangle$. Fig.~\ref{fig:borncartesianfull}(c)  and (d) show the corresponding plots for $U/t=4$. We see that the $U/t=4$ data clearly shows the formation of a flat mode leading to a narrow subgap weight at $\qq=[0,0]$ at a finite frequency. The upper edge of the collective modes at $U/t=3$ are too close to the continuum to see this clearly. Here, we would like to note that the dispersion of the large momentum, high frequency collective mode depends on whether one uses a ``Cartesian" representation of fluctuations (as done here) or works with the amplitude-phase representation. In the amplitude-phase representation, the collective mode frequencies are a bit lower and better separated from the two-particle continuum.  There is also pileup of low energy weight observed near the $M$ point. This shows that this simple approximation is able to capture the occurence of subgap two-particle spectral weight at long wavelengths in a weakly disordered superconductor.
In the next section, we will work in the amplitude-phase co-ordinates and further simplify our model
to obtain an analytic understanding of the systematic trends in the two-particle spectral functions at weak disorder.

\section{CDW Fluctuations and Effective 2 band Model}
\label{sec4}

In the previous section, we have converted the problem of attractive fermions in the presence of a disorder
potential to that of bosonic collective fluctuations of the superconducting order parameter scattered by an
effective disorder in pairing amplitudes and local potentials. We have also seen that a Born approximation
calculation using the variance and covariance of the effective disorder fields obtained from BdG solutions
reproduce the basic phenomena of a narrow subgap weight below the continuum threshold at $\qq=[0,0]$.
In order to make further analytic progress, we need an analytic handle on the static fluctuation correlators,
$\langle|\delta v(\qq)|^2 \rangle, \langle|\delta \Delta(\qq)|^2 \rangle$ and $\langle\delta v(\qq)\delta\Delta(-\qq)\rangle$.

We consider a system with $U/t=3$ and $n=0.875$ and plot the correlators $\langle|\delta v(\qq)|^2 \rangle$,
$\langle|\delta \Delta(\qq)|^2 \rangle$ and $\langle\delta v(\qq)\delta\Delta(-\qq)\rangle$, calculated from the
spatially inhomogeneous mean-field solutions. Fig.~\ref{fig:correlator1}(a)-(c) show the color-plot of the
effective potential correlator $\langle|\delta v(\qq)|^2\rangle$ as a function of $\qq$ for increasing disorder
strength $V/t=0.1,~0.5,~1.0$ respectively. Fig.~\ref{fig:correlator1}(d)-(f) show the corresponding plots
for $\langle|\delta \Delta(\qq)|^2\rangle$, and Fig.~\ref{fig:correlator1}(g)-(i) show the cross-correlator
between the pairing amplitude and the effective potential. Two interesting trends can be seen in these plots:
(a) $\langle|\delta v(\qq)|^2\rangle \gg \langle|\delta\Delta(\qq)|^2\rangle, \langle\delta v(\qq)\delta\Delta(-\qq)\rangle$,
so that it is reasonable to only consider the effects of effective static potential fluctuations and neglect the other
correlators. We have specifically checked that keeping the other correlators finite does not change the qualitative
understanding we get from this simplified assumption. (b) We immediately notice that the correlations are peaked
at $\QQ=[\pi,\pi]$ for weak disorder. As we increase the disorder strength, the peak at $\QQ$ broadens
(See Fig.~\ref{fig:correlator1}(c) and (f)), while also increasing in strength. The strong peak at $\QQ=[\pi,\pi]$
is due to proximity to the charge density wave (CDW) instability~\cite{auerbach2012interacting} of this model at half-filling
at the commensurate wave-vector. Although we are away from half-filling, and hence do not have a static
CDW order ($\langle \delta v(\QQ)\rangle=0$), the broken translation invariance due to disorder
creates strong spatial fluctuations with wave-vector $\QQ$.  

\begin{figure*}
\includegraphics[width=0.7\textwidth]{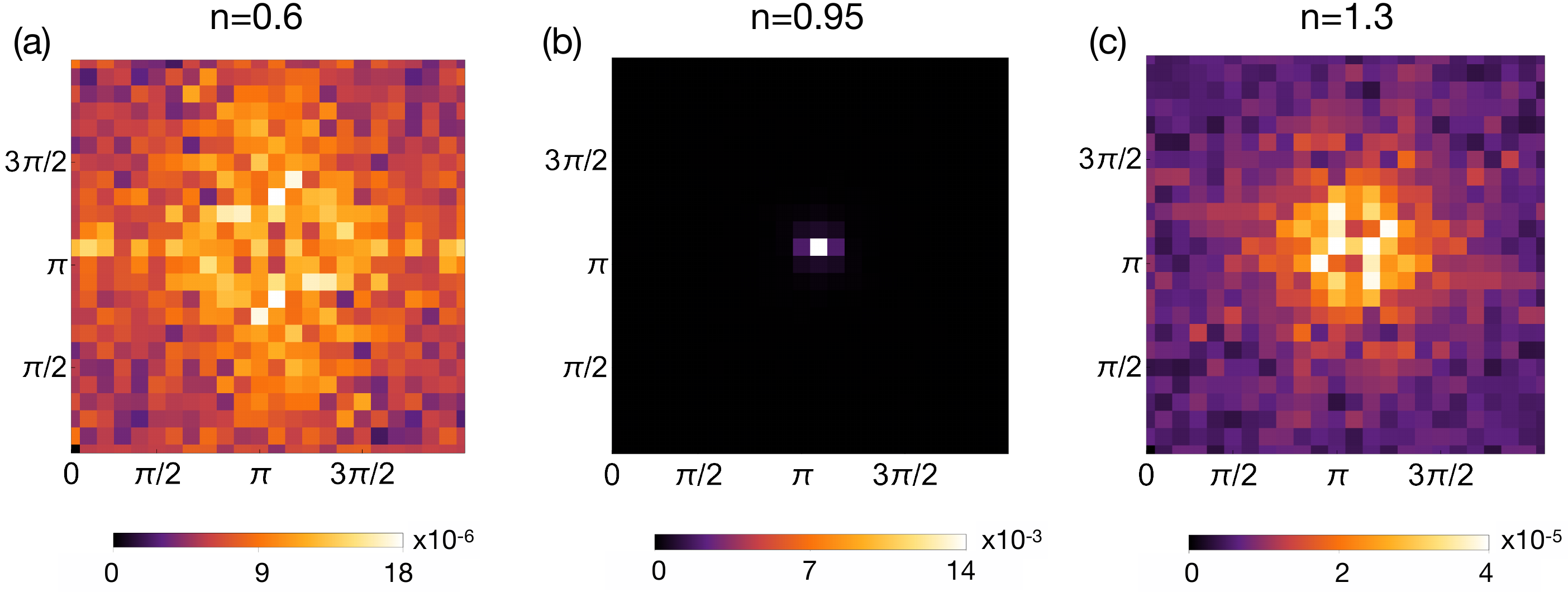}
\caption{Color-plot of the disorder averaged static fluctuations of the Hartree field $\langle |\delta v(\qq)|^2\rangle$ over the $2D$ Brillouin zone for (a) $n=0.6$, (b) $n=0.95$ and (c) $n=1.3$. We see that the correlator is strongly peaked at $\qq=[\pi,\pi]$ near half-filling $(n=1)$ due to proximity to the CDW instability and becomes more diffuse as we deviate from half-filling on either side. The correlators are calculated from the BdG solutions with $U/t=3$ and $V/t=0.1$ and averaged over $15$ disorder realizations. }
\label{fig:correlator2}
\end{figure*}

We can tune the system away from the CDW instability by changing the average density away from half-filling. In Fig.~\ref{fig:correlator2}(a)-(c), we plot
$\langle |\delta v(\qq)|^2\rangle$ for a system at a fixed weak
disorder ($V/t=0.1$), but with different average densities across half-filling
($n=0.6$ for (a), $n=0.95$ for (b), and $n=1.3$ for (c)). We clearly see that as we approach
half-filling, the correlator peaks at $\QQ$, while the weight is more diffusely spread over the Brillouin zone
as we move away from it. This reinforces the idea that the strong peak at $\QQ$ is a signature of the nearby
CDW instability in the system. 

The strong peak of the potential fluctuations around $\QQ$ motivates a simpler model where
only $\delta v(\QQ)$ is considered and all other static fluctuations are neglected. In this case, the
mode at $\qq$ is coupled to the mode at $\qq +\QQ$, and one can work within a mode coupling theory in a ``magnetic Brillouin zone''
corresponding to the commensurate wave-vector with a doubling of the degrees of freedom. Further, in this
case, we will work with the amplitude-phase co-ordinates for the quantum fluctuation of the superconducting
order-parameter, i.e. expand $\Delta(r,\tau)=[\Delta_0+\delta \Delta(r)+\eta(r,\tau)]e^{i\theta(r,\tau)}$,
where $\eta$ and $\theta$ are the amplitude and phase of the quantum fluctuations. This allows us to
consider the nature of the subgap mode and the partitioning of the subgap weight into the amplitude
and phase degrees of freedom cleanly. Within this approximation, the fluctuation action can now be written as
\begin{widetext}
\beq
S_{fl}=\frac{1}{2}\sum_{\qq,\omega_m} [\Gamma(\qq,i\omega_m),\Gamma(\qq-\QQ,i\omega_m)] \left[ \begin{array}{cc}
                                                                                                   \hat{D}^{-1}(\qq,i\omega_m) & \delta v(\QQ) \hat{F}(\qq,i\omega_m)\\
                                                                                                   \delta v(\QQ)^\ast \hat{F}(\qq+\QQ,i \omega_m) &\hat{D}^{-1}(\qq+\QQ,i\omega_m)
                                                                                                 \end{array}\right] \left[ \begin{array}{c} \Gamma(-\qq,-i\omega_m)\\
                                                                                                                   \Gamma(-\qq+\QQ,-i\omega_m)        \end{array}\right]
\eeq
where $\Gamma(\qq,i\omega_m)=[\eta(\qq,i\omega_m),\theta(\qq,i\omega_m),\xi(\qq,i\omega_m)$ is a
three-component field containing the amplitude ($\eta$), the phase ($\theta$) and the Hartree potential ($\xi$) fluctuations. In some cases, we will suppress the density fluctuations and work with a two-component $\Gamma$ field. Here $\hat{D}^{-1}(\qq,i\omega_m)$ is the inverse propagator of the quantum fluctuations in
 the translation invariant system in the amplitude-phase-potential co-ordinates,
\bqa
\no D^{-1}_{11}(\qq,i\omega_m)&=& \frac{1}{U} + \frac12 \sum_{\kk,i\omega_n}\text{Tr }^N G(\kk+\qq,i\omega_n+i\omega_m)\sigma^1G(\kk,i\omega_n)\sigma^1\\
\no D^{-1}_{12}(\qq,i\omega_m)&=& \frac{i}{4}(i\omega_m)\sum_{\kk,i\omega_n}\text{Tr }^N G(\kk+\qq,i\omega_n+i\omega_m)\sigma^1G(\kk,i\omega_n)\sigma^3 = -D^{-1}_{12}(\qq,i\omega_m)\\
\no D^{-1}_{22}(\qq,i\omega_m)&=& (i\omega_m)^2 \kappa(\qq,i\omega_m)+\sum_{\hat\delta=\pm \hat x,\hat y}(1-\cos(\qq\cdot \hat\delta)\Theta_\delta+\chi(\qq,i\omega_m)\\
\no D^{-1}_{13}(\qq,i\omega_m) &=& \frac12 \sum_{\kk,i\omega_n}\text{Tr }^N G(\kk+\qq,i\omega_n+i\omega_m)\sigma^3G(\kk,i\omega_n)\sigma^1 = D^{-1}_{31}(\qq,i\omega_m)\\
\no D^{-1}_{23}(\qq,i\omega_m)&=&-\frac{i}{4}(i\omega_m)\sum_{\kk,i\omega_n}\text{Tr }^N G(\kk+\qq,i\omega_n+i\omega_m)\sigma^3G(\kk,i\omega_n)\sigma^3=-D^{-1}_{32}(\qq,i\omega_m)\\
 D^{-1}_{33}(\qq,i\omega_m)& =& \frac{1}{U}+\frac12 \sum_{\kk,i\omega_n}\text{Tr }^N G(\kk+\qq,i\omega_n+i\omega_m)\sigma^3G(\kk,i\omega_n)\sigma^3
\eqa
where $\kappa$ is the generalized compressibility $\Theta_\delta$ is the kinetic energy, and $\chi$ is the current-current correlator in the system given by
\bqa
\no \kappa(\qq,i\omega_m) &=& \frac18 \sum_{\kk,i\omega_n}\text{Tr }^N G(\kk+\qq,i\omega_n+i\omega_m)\sigma^3G(\kk,i\omega_n)\sigma^3\\
\no \Theta_\delta &=& \frac{t}{2}\sum_{E_\kk >0}v_\kk^2\cos(\kk\cdot \hat\delta)\\
 \chi(\qq,i\omega_m) &=& \frac{t^2}{8}\sum_\kk (\epsilon_{\kk+\qq} - \epsilon_\kk)^2 \sum_{i\omega_n}\text{Tr }^N G(\kk+\qq,i\omega_n+i\omega_m)\sigma^0G(\kk,i\omega_n)\sigma^0
\eqa
The details of the $D^{-1}$ matrix is derived in Appendix~\ref{app:2band}. The off-diagonal scattering matrix $\hat{F}$,
which gives the amplitude to scatter between $\qq$ and $\qq+\QQ$ modes, are given by
\begin{gather}
\no F_{11}=\frac12 \Pi_{113}\hphantom{3333}F_{21}=-\frac{i}{4}  (i\omega_m) \Pi_{313} \hphantom{3333}  F_{12}=\frac{i}{4}  (i\omega_m) \Pi_{133}\hphantom{3333}F_{33}=\frac{1}{2}\Pi_{333}\\
    \no F_{13}=\frac12\Pi_{133}\hphantom{3333}F_{23}=-\frac{i}{4} (i\omega_m)\Pi_{333}\hphantom{3333}F_{31}=-\frac{1}{2}\Pi_{313}\hphantom{3333} F_{32}=-\frac{i}{4}(i\omega_m)\Pi_{333}\\
F_{22}= \frac{(i\omega_m)^2}{8}\Pi_{333}+\frac{1}{8}f_{22}^0 - \left[\frac{(\epsilon_{\kk+\qq}-\epsilon_\kk)(\epsilon_\kk-\epsilon_{\kk+\qq+\QQ})}{8}\right]\Pi_{003} 
\end{gather}
with 
\begin{eqnarray}
   \no \left[ f(\kk,\qq)\right] \Pi_{abc}&=& \sum_{\kk,i\omega_n}f(\kk,\qq) \text{Tr }^NG(\kk+\qq,i\omega_n+i\omega_m) \sigma^a G(\kk,i\omega_n)\sigma^b G(\kk+\qq+\QQ,i\omega_n+i\omega_m) \sigma^c\\
    f_{22}^0&=&\sum_{\kk,i\omega_n} (\epsilon_{\kk+\qq}-\epsilon_{\kk-\qq}) \text{Tr }^N G(\kk,i\omega_n)\sigma^zG(\kk+\QQ,i\omega_n)\sigma^z
\end{eqnarray}
\end{widetext}
where $\text{Tr }^N$ is the trace over Nambu indices, and $\sigma^0$ is the identity matrix. One can now invert the full $6\times 6$  inverse propagator matrix to obtain the Green's functions, 
and then construct the spectral functions of the collective bosons. Note that if we are interested in the spectral 
function of the fluctuations at a fixed $\qq$ (as opposed to the matrix element to scatter from $\qq$ to $\qq+\QQ$), 
the answers only involve $|\delta v(\QQ)|^2$. We can then replace $|\delta v(\QQ)|^2$ by its disorder average 
$\langle |\delta v(\QQ)|^2\rangle$, and consider the problem as a function of this single parameter. 
This simplified model, where the effects of disorder has been 
reduced to a single parameter, contains all the physics behind the systematic changes of the two -particle spectral functions at weak disorder.
\begin{figure*}
\includegraphics[width=0.95\textwidth]{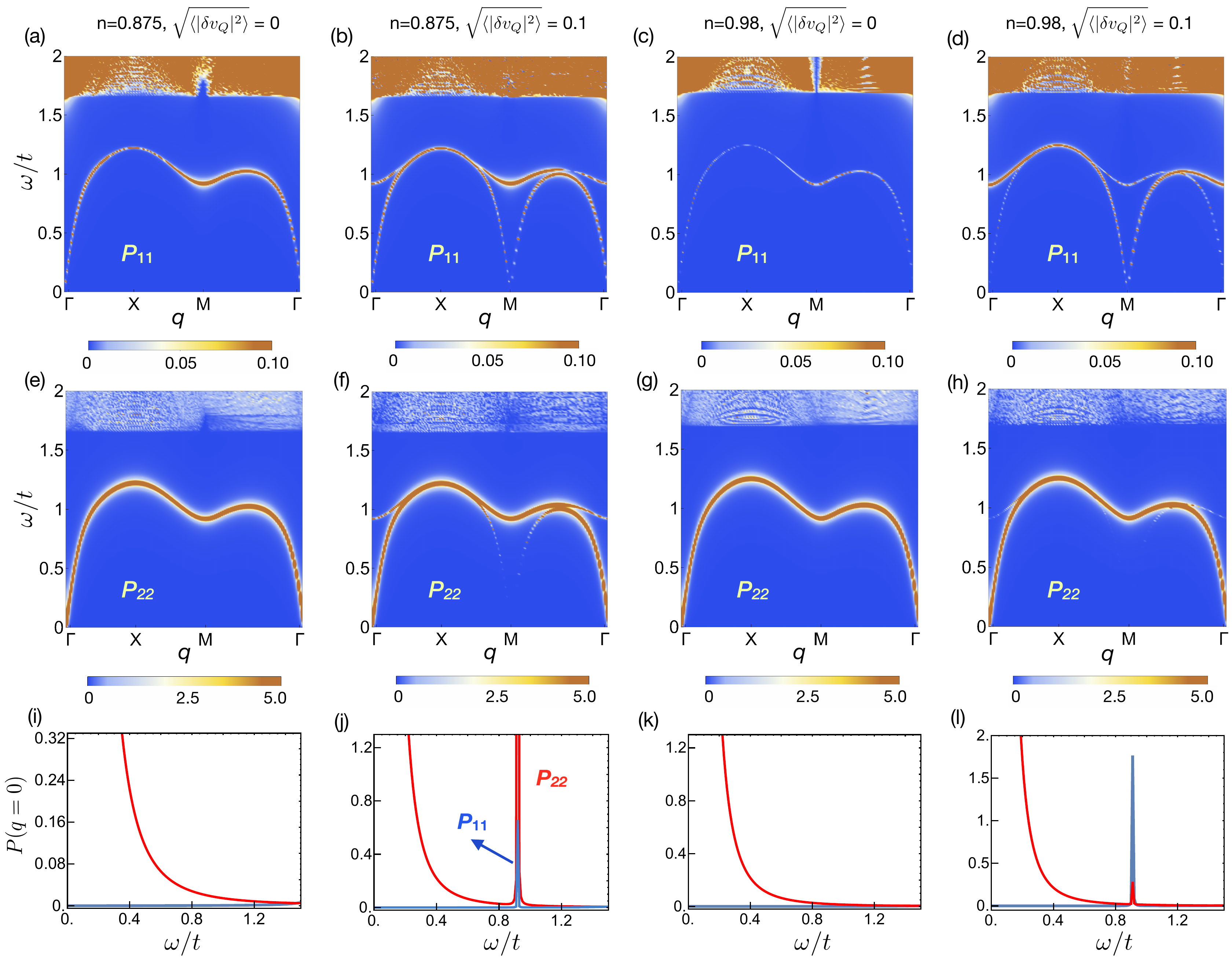}
\caption{Results from the mode coupling theory. (a)-(b): The amplitude spectral functions $P_{11}$ for $n=0.875$ (a) corresponds to clean case $\sqrt{\langle |\delta v(\QQ)|^2\rangle}=0$ while (b) corresponds to a weak disorder with $\sqrt{\langle| \delta v(\QQ)|^2\rangle}=0.1t$. (c) and (d) are similar to (a) and (b) respectively but for $n=0.98$. The subgap mode is clearly visible for the disordered case for both (b) $n=0.875$ and (d) $n=0.98$. The phase spectral functions $P_{22}$ corresponding to the cases in (a)-(d) are plotted in (e)-(h) respectively. The subgap mode is clearly seen at (f) $n=0.875$ while it is strongly supressed at (h) $n=0.98$. Moreover the phase channel is supressed close to half-filling. The EDC at $\qq=[0,0]$ for the amplitude (blue) and phase (red) spectral functions for the cases corresponding to (a)-(d) are plotted in (i)-(l) respectively. All the plots have a low energy tail from the pure phase Goldstone mode. In the disordered cases (j) and (l), the subgap mode is clearly seen. At $n=0.98$ close to half-filling the subgap weight is dominated by the amplitude component while at $n=0.875$ away from half-filling the phase component dominates. The calculations are done on a $100\times 100$ lattice. }
\label{fig:twobandampphase}
\end{figure*}
\subsection{Collective Modes with Pairing Fluctuations}
 
 We first apply our mode coupling model to investigate the collective modes solely in presence of dynamic pairing fluctuations, i.e. we set the density fields, $\xi(r.\tau)=0$. In Fig.~\ref{fig:twobandampphase}(a) and (b) we plot the amplitude spectral function obtained from the mode coupling theory for a system at $n=0.875$ and $U/t=3$. Fig.~\ref{fig:twobandampphase}(a) corresponds to the clean case, i.e. $\langle |\delta v(\QQ)|^2\rangle=0$, while Fig.~\ref{fig:twobandampphase}(b) corresponds to $\langle |\delta v(\QQ)|^2\rangle=0.01 t^2$. The value of $\langle |\delta v(\QQ)|^2\rangle$ is chosen to be in a realistic regime for systems with weak disorder. In Fig.~\ref{fig:twobandampphase}(a), we see the standard collective modes in the homogeneous system, which disperses linearly at low $\qq$. The amplitude weight in this mode goes to $0$ as $\qq \rightarrow 0$. In contrast, 
 Fig.~\ref{fig:twobandampphase}(b) clearly shows the almost non-dispersive weight at finite subgap energy. In the long wavelength limit, we have a coupling between the zero energy pure phase Goldstone mode at $\qq=[0,0]$  with the collective mode at $\QQ$ (with energy $\omega_{\QQ}$), which has both amplitude and phase components. 
 At weak disorder, the off-diagonal coupling $\delta v_\QQ F \ll \omega_\QQ$, and hence the finite frequency spectral weight appears at $\omega_0 \sim \omega_\QQ - (\delta v_\QQ F)^2/2 \omega_\QQ \simeq \omega_\QQ$. The strong $[\pi,\pi]$ scattering due to the disorder thus creates the narrow subgap weight at $\qq=[0,0]$. Note that as a consequence a mirror image of the mode near $[0,0]$ shows up around $[\pi,\pi]$, which is seen as a pileup in the low energy spectral weight around the $M$ point. Fig.~\ref{fig:twobandampphase}(e) and (f) show the phase spectral functions corresponding to the amplitude spectral functions shown in Fig.~\ref{fig:twobandampphase}(a) and (b) respectively. We note that the subgap mode in this case, shows up both in the amplitude and phase spectral functions. Fig.~\ref{fig:twobandampphase}(i) and (j) show the energy distribution curves at $\qq=[0,0]$ for the clean and the disordered cases respectively. In the clean case [Fig.~\ref{fig:twobandampphase}(i)], one can clearly see that there is no subgap Higgs weight, while the low energy phase weight gets contribution from the Goldstone mode. In contrast, in Fig.~\ref{fig:twobandampphase}(j), the disorder scattering  creates additional spectral weight in the amplitude and phase channels at approximately the energy of the homogeneous collective mode at $\QQ$. This weight is narrowly distributed in energy and is well separated from the low energy collective mode weight. We note that in a real disordered system, the disorder scattering happens with all momentum transfers, with the scattering amplitude peaking at $\QQ$. In this case, one would expect the spectral weight to be smeared over a larger energy window. Further, as the scattering becomes diffuse with increasing disorder, one would expect this mode to broaden, which is what is seen in the numerics around the disordered inhomogeneous BdG saddle point.
\begin{figure}
	\includegraphics[width=0.49\textwidth]{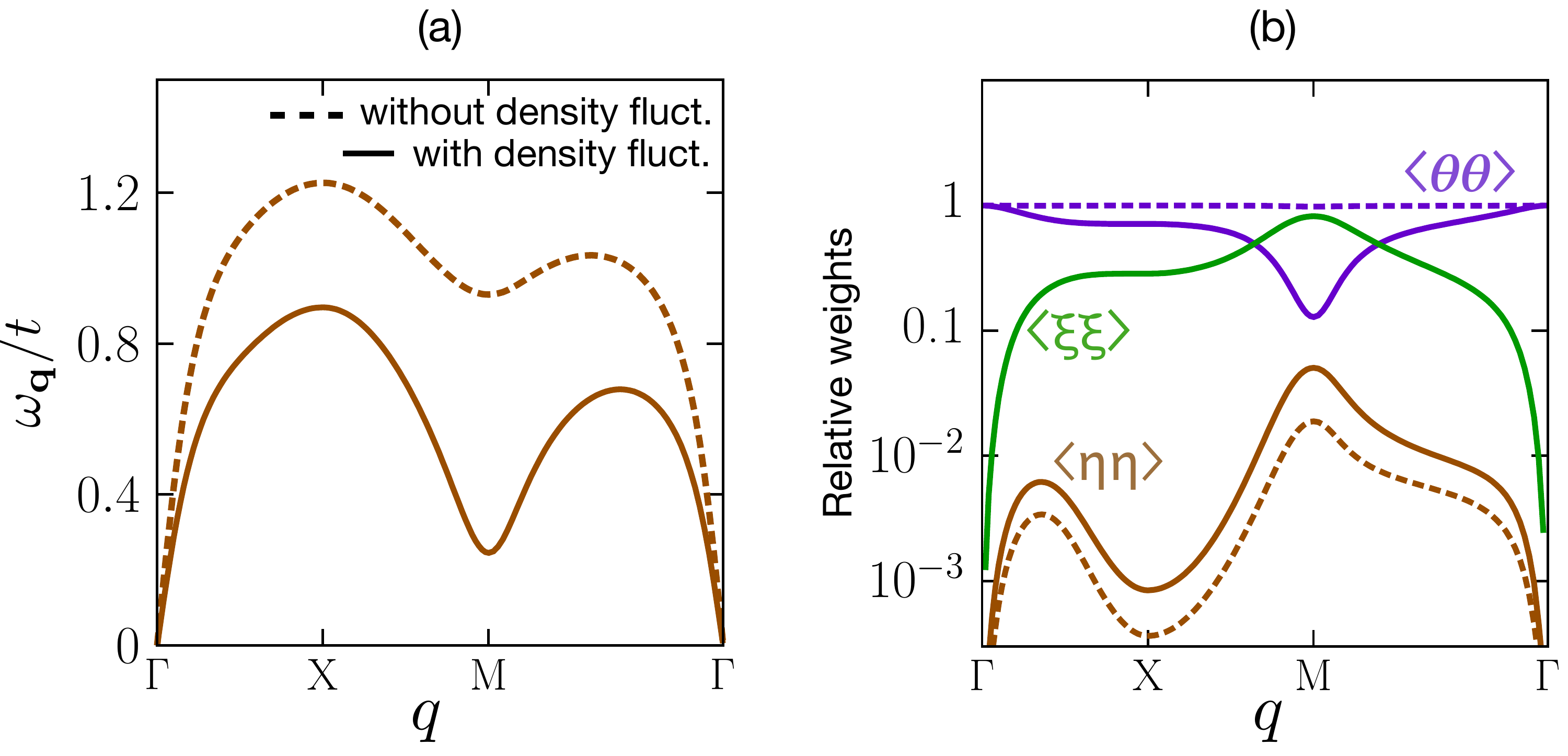}
	\caption{(a) The collective mode dispersion of a clean $s$-wave supercondcutor with (solid lines) and without (dashed lines) density fluctuation at $U/t=3$ and $n=0.875$. The inclusion of density fluctuations lowers the collective mode frequency near the $M$ point substantially. (b) The relative weight of the amplitude (brown), phase (purple) and density (green) fluctuations in the collective modes are shown. Near the M point there is a transfer of weight from the phase to the density channel.}
	\label{fig:collectivedenfluct}
\end{figure}

We now focus on the amplitude and phase components of the spectral weight at $\qq=[0,0]$ at finite frequency.
In the numerical results, we have seen that as we move away from half-filling the phase component of the spectral weight increases at the expense of the amplitude. To understand this trend, we consider the scattering matrix $\hat{F}(\QQ,i\omega_m)$ at the particle-hole symmetric half-filling limit, where the chemical potential $\mu-v_0=0$. Using the fact that $\epsilon_{\kk+\QQ}=-\epsilon_\kk$, and ${G}(\kk+\QQ)=\sigma^1G(\kk)\sigma^1$, one can show that
\bqa
\no F_{11}(\QQ,i\omega_m)&=&\sum_\kk \frac{\epsilon_\kk}{E_\kk}\frac{1}{(i\omega_m)^2-4E_\kk^2}=0\\
 F_{12}(\QQ,i\omega_m)&=&-\frac{i}{2}(i\omega_m)\sum_\kk \frac{\Delta}{E_\kk}\frac{1}{(i\omega_m)^2-4E_\kk^2}\\
\no F_{22}(\QQ,i\omega_m)&=&-\frac{(i\omega_m)^2}{4}\sum_\kk \frac{\epsilon_\kk}{E_\kk}\frac{1}{(i\omega_m)^2-4E_\kk^2}=0.
\eqa 
Here $F_{21}(\QQ,i\omega_m)=-F_{12}(\QQ,i\omega_m)$. We note that the $\qq=[0,0]$ mode is always a pure phase mode. Since particle-hole symmetry prohibits a coupling between the phase component at $\qq=[0,0]$ with the phase component at $\qq=[\pi,\pi]$ ($F_{22}=0$) at half-filling, the Goldstone mode primarily couples to the amplitude component of the  $[\pi,\pi]$ mode near half-filling. Hence, in the weak disorder limit, when the off-diagonal coupling between the modes are much smaller that $\omega_\QQ$, the finite frequency weight is mostly in the amplitude channel. To see this, we plot the two-particle spectral function for a system with $n=0.98$ close to half-filling at a weak disorder of $\langle |\delta v(\QQ)|^2\rangle=0.01 t^2$ in Fig.~\ref{fig:twobandampphase}(d) (amplitude spectral function) and Fig.~\ref{fig:twobandampphase}(h) (phase spectral function) respectively. The corresponding clean case spectral functions are shown in Fig.~\ref{fig:twobandampphase}(c) and (g) respectively. In this case we can clearly see that at $\qq=[0,0]$, the finite frequency subgap spectral weight is seen mostly in the amplitude channel, while the phase spectral weight is concentrated near $\omega=0$. This is clearly illustrated in the energy distribution cuts (at $\qq=[0,0]$ in Fig.~\ref{fig:twobandampphase}(k) (clean case) and Fig.~\ref{fig:twobandampphase}(l) (at weak disorder). In Fig.~\ref{fig:twobandampphase}(l), we clearly see a finite subgap weight dominated by the Higgs channel, which is spectrally separated from the low energy weight in the phase channel. As one moves away from the half-filling, $F_{22}$ increases, and hence the phase component of the finite frequency spectral weight increases, as seen in the more accurate numerical calculations shown in Section II.

We note that while our simplified toy model of mode coupling correctly predicts the trends, it does not provide quantitatively correct answers; e.g. at $n=0.875$, the numerical calculations show a preponderance of amplitude spectral weight, while we need to go much closer to half-filling ($n=0.98$) to see this. Thus the mode coupling theory should be used to understand systematic trends and should not be used to directly compare quantitatively with the numerical results. However, it still provides valuable insights behind the systematic trends, which is hard to obtain from more sophisticated calculations.

\subsection{Effects of Density Fluctuations}

We have so far considered the quantum fluctuations in the superconducting order parameter to determine the collective mode spectrum, while the Hartree field was accounted for only through its static fluctuations; i.e. the standard deviation of its spatial variations in the inhomogeneous mean-field solutions. We now consider the effects of dynamic density fluctuations on the two-particle spectral weight by considering
\begin{align*}
\xi(r)\to \xi_0 +\delta v(r)+\xi(r,\tau)
\end{align*}
where $\xi$ represent the temporally and spatially varying particle-hole fluctuations. We note that this is a key difference between approximations made in Ref.~\onlinecite{HiggsAbhisek} and Ref~\onlinecite{Benfatto1}, which reach different conclusions on the exact location of the subgap weight and its amplitude-phase distribution. Compared to Ref.~\onlinecite{HiggsAbhisek}, Ref.~\onlinecite{Benfatto1} finds broad subgap weights at lower frequencies, with much larger phase spectral weight. Before we consider the mode coupling theory due to disorder scattering, we first consider how the collective mode in the uniform system changes due to inclusion of these density fluctuations; i.e. set $\langle |\delta v(\QQ)|^2\rangle=0$.

In this case, $\Gamma$ is a three-component vector, and the collective modes can be found from the zeroes of the determinant of the $3\times 3$ matrix that forms the inverse propagator for the fluctuations. In Fig.~\ref{fig:collectivedenfluct}(a), we plot the dispersion of the collective modes in a clean system (at $U/t=3$, $n=0.875$) calculated with (solid line) and without (dashed line) considering the density fluctuations. The main effect of including the density fluctuations is to lower the collective mode frequency at large $\qq$, especially around the $M$ point. This is once again due to the proximity of the CDW instability. At the CDW instability, one expects the collective mode frequency to go to $0$ at the $M$ point. We note that one can integrate out the density fluctuations to obtain an effective amplitude-phase correlator, but the collective mode dispersion remains essentially same whether one works with a $3\times 3$ propagator or an effective $2 \times 2$ propagator. In Fig.~\ref{fig:collectivedenfluct}(b), we plot the relative weights of the amplitude, phase and density sectors in the collective modes (by considering the eigenvector which gives the collective mode). We see that the weight of the amplitude sector is almost unaffected by inclusion of the density fluctuations. In the long wavelength limit, the weight of the density fluctuations go to zero. The main effect of the density fluctuations can be seen near $M=[\pi,\pi]$, where the weight is transferred from the phase to the density channel.

\begin{figure}
	\includegraphics[width=0.49\textwidth]{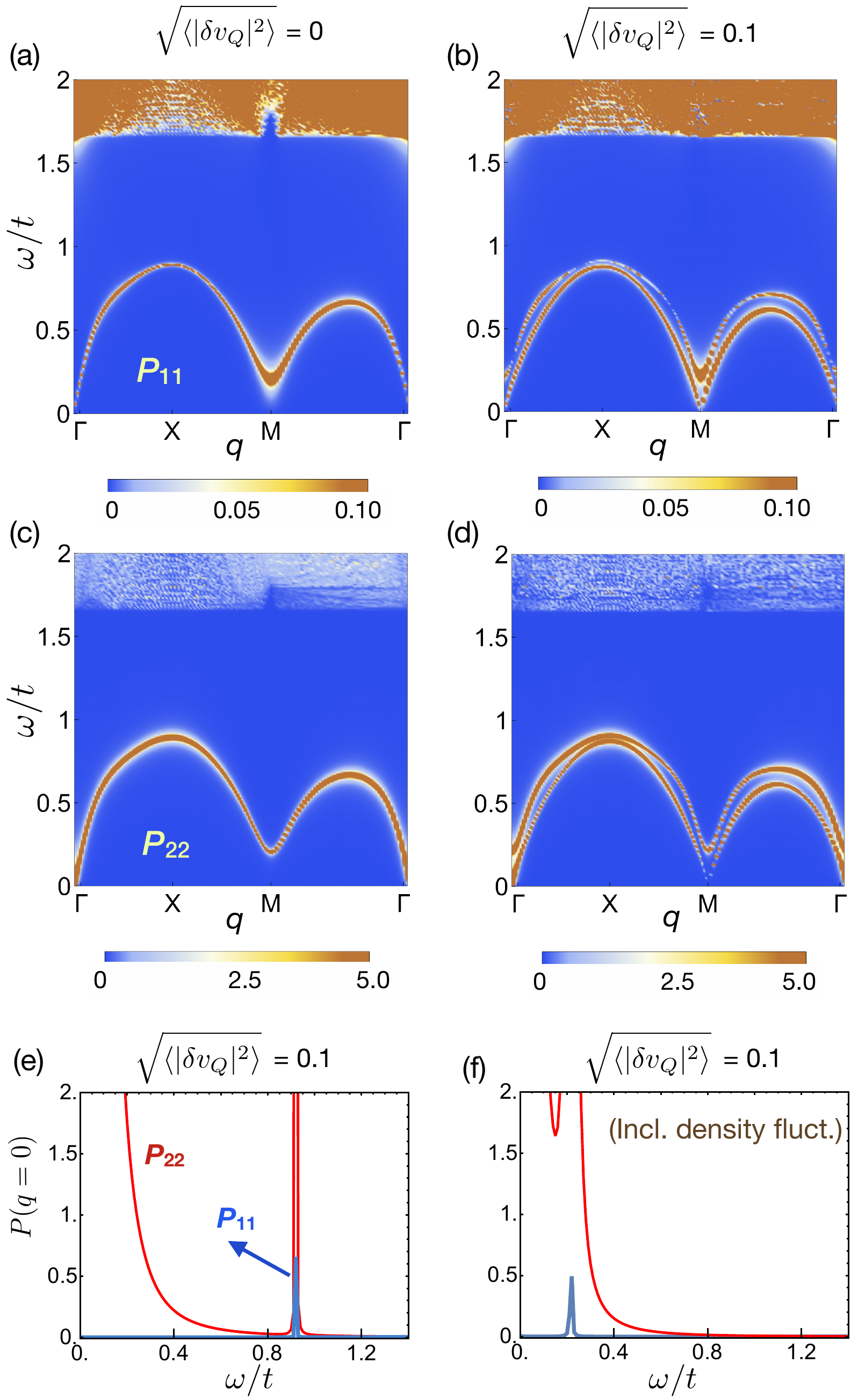}
	\caption{ (a) and (b) the amplitude spectral fluctuations and (c) and (d) the phase spectral functions calculated using the mode coupling theory including the effects of density fluctuations. (a) and (c) correspond to the clean case while (b) and (d) correspond to the weak disorder with $\sqrt{\langle|\delta v(\QQ)|^2\rangle}=0.1t$. (e) and (f) show the EDC at $\qq=[0,0]$ both (e) without and (f) with density fluctuations at a weak disorder of $\sqrt{\langle|\delta v(\QQ)|^2\rangle}=0.1t$. Including density fluctuations causes the subgap mode at $\qq=[0,0]$ to shift to a lower frequency, and increases the relative contribution of the phase mode. The calculations were done within mode coupling theory at $U/t=3$ and $n=0.875$ on a $100\times 100$ lattice. }
	\label{fig:bornampphaseden}
\end{figure}

We now consider the effects of density fluctuations on the spectral functions in the disordered system through the mode-coupling theory. In Fig.~\ref{fig:bornampphaseden}(a) and (b) we plot the amplitude spectral function of a system at $U/t=3$ and $n=0.875$  for the clean case ($\langle |\delta v(\qq)|^2\rangle=0$) and for a weak disorder ($\langle| \delta v(\qq)|^2\rangle=0.1$) respectively. Fig.~\ref{fig:bornampphaseden}(c) and (d) show the corresponding phase spectral functions. While the amplitude spectral function is almost unchanged, we clearly see two split bands in the phase spectral function. Fig.~\ref{fig:bornampphaseden}(e) shows the spectral function at $\qq=[0,0]$ as a function of energy for the disordered system in absence of dynamic density fluctuations, while Fig.~\ref{fig:bornampphaseden}(f) shows the same quantity when these fluctuations are included. Two trends are clearly seen: (i) The additional feature at finite frequencies is pushed down when density fluctuations are included. Within the mode coupling theory, as the collective mode frequency at $\QQ$ comes down, it pushes the additional feature at $\qq=[0,0]$ downwards. Note that a sharp additional feature can be seen in this case with identifiable amplitude and phase contributions, although it is no longer spectrally separated from the tail of the spectral weight from the zero energy Goldstone mode. In a theory where scattering at all momenta are kept, this feature will be broadened further. (ii) The additional feature has much larger phase component compared to the case without density fluctuations. We note that since the energy of the mode at $\QQ$ is smaller in this case, there is a larger mixing between the $\qq=[0,0]$ Goldstone mode (which is a pure phase mode) and the mode at $[\pi,\pi]$, leading to a larger phase component in the subgap weight. Thus our two-mode model is able to accurately capture this trend and can also resolve the discrepancies between  Ref.~\onlinecite{HiggsAbhisek} and Ref.~\onlinecite{Benfatto1}.

\section{Conclusion}
\label{sec5}
In this paper we have shown that the collective modes and two-particle spectral weight of a weakly disordered superconductor can be obtained from a model where the translation invariant collective modes are scattered by an effective disorder. Starting with a microscopic theory of attractive fermions in presence of random potential disorder, we construct this effective theory by expanding the action around a translation invariant saddle point in both the static spatial fluctuations induced by disorder and the dynamic quantum fluctuations. We can thus construct the parameters of this effective model starting from a fermionic theory. We show that a simple Born approximation for the disorder scattering of the collective modes reproduces the long wavelength ($\qq=[0,0]$) subgap spectral weight at finite frequencies, which have been seen earlier in numerical calculations around the inhomogeneous BdG mean field solution.

In the attractive Hubbard model on a square lattice, the system undergoes a CDW instability with $\QQ=[\pi,\pi]$ at half-filling. In presence of disorder, there are strong static fluctuations of density and pairing fields at this commensurate wave-vector, even when the system is away from half-filling. Such strong fluctuations will be a generic feature of systems near a CDW instability. The effective disorder seen by the collective modes thus shows a strong peak at this wave-vector. This leads to a simple mode coupling theory (coupled by random static fluctuations), which provides analytic insight into the subgap weight in the two-particle spectral functions at $\qq=[0,0]$ and captures the trends seen in the numerical calculations. The subgap weight is formed by disorder scattering of the $[\pi,\pi]$ mode, and appears around the energy of this collective mode. The particle-hole symmetry at half-filling ensures that this mode consists primarily of amplitude fluctuations near half-filling. As we move away in density, the phase contribution to this mode increases. Including dynamic density fluctuations lowers the frequency of the $[\pi,\pi]$ mode substantially, and hence the spectral separation between the tail of the Goldstone weight and this mode is lost. The lower frequency also implies a larger mixing of the  $\qq=[0,0]$ Goldstone mode (which is a pure phase mode),  and hence phase contribution to the subgap weight increases substantially in this case.

We have thus obtained a generic framework to obtain two-particle spectral weight of weakly disordered superconductors. Therefore close to a CDW transition, we obtain a simpler mode coupling theory. The trends explain the discrepancies between Ref.~\onlinecite{HiggsAbhisek} and Ref.~\onlinecite{Benfatto1}. The extension of this framework to stronger disorder strengths by going beyond the simple Born approximation is left for future work.

\begin{acknowledgements}
P.P.P and R.S. acknowledge the NIUS program for seeding their collaboration on this work. R.S. acknowledges support of the Department of Atomic Energy, Government of India, under Project Identification No. RTI 4002. The computations were performed using the computational facilities at the Department of Theoretical Physics, TIFR Mumbai and at Physics Department, Technion. 
\end{acknowledgements}
\appendix
\begin{widetext}

\section{Fluctuations in Cartesian Coordinates}
\label{app:cartesian}
 We work with the disordered $2D$ attractive Hubbard model on a square lattice. The Hamiltonian is given by,
\begin{align}
    H&= -t \sum_{\langle rr'\rangle\sigma}c^\dagger_{r\sigma}c_{r'\sigma} - U \sum_r n_{r\uparrow}n_{r\downarrow} +\sum_r (v_r - \mu) n_r
\end{align}
where $c^{\dagger}_{r\sigma}(c_{r\sigma})$ is the creation (annihilation) operator for an electron with spin $\sigma$ on site $r$, and $\mu$ is the chemical potential. Here $t$ is the nearest neighbour hopping parameter, and $U$ is the local attractive interaction between the electrons. $v_r$ is an independent random variable for each site that is uniformly sampled from $[-V/2,V/2]$. We work in the imaginary time path integral formalism and decouple the attractive interaction in the density channel and the pairing  channel to get the following action 
\begin{align}
    S &= \int_0^\beta d\tau \sum_r \frac{|\Delta(r,\tau)|^2+ |\xi(r,\tau)|^2}{U} - \int d\tau d\tau' \sum_{rr'}\psi^\dagger(r,\tau) \mathcal{G}^{-1}(r\tau, r'\tau') \psi(r',\tau')\\
    \mathcal{G}^{-1} &= \delta(\tau-\tau')\begin{pmatrix}\left(\partial_\tau +\mu-v^{eff}(r,\tau)\right)\delta_{rr'}+t\delta_{rr'} & -\Delta(r,\tau)\delta_{rr'} \\ -\Delta^*(r,\tau)\delta_{rr'} & -(\partial_\tau +\mu- v^{eff}(r,\tau))\delta_{rr'}-t\delta_{rr'}\end{pmatrix}\\
\no  \textrm{where} ~~~  v^{eff}(r,\tau)&= v(r) -\xi(r,\tau)~.
\end{align}
Here $\psi^\dagger(r,\tau)=\{\bar{f}_\uparrow(r,\tau),f_\downarrow(r,\tau)\}$ is the Grassman Nambu spinor for the fermionic operators. $\xi$ is the Hartree shift and $\Delta$ is the superconducting order parameter (s-wave). Solving the mean-field equations gives us the static mean-field values for $\Delta(r)$ and $\xi(r)$~\cite{HiggsAbhisek}. All the calculations are done at zero temperature. 

Considering the static and dynamic fluctuations of the pairing field, we make the substitution $\Delta(r,\tau) \to \Delta_0+\delta\Delta(r) +\lambda(r,\tau)$, where $\delta\Delta(r)$ is the spatial fluctuation of the mean-field order parameter about its average value, with $\langle\delta\Delta(r)\rangle_r=0$, where we average over the spatial position $r$. Similarly, we write $v^{eff}(r,\tau)=v_0+\delta v(r)+\xi(r,\tau)$. Note that $\Delta_0$ and $v_0$ represent the mean-field solutions in the clean superconductor. The Green's function becomes 
\begin{align}
    \mathcal{G}^{-1}(\kk+\qq,\kk,i\omega_m+i\omega_n,i\omega_n)=G^{-1}(\kk,i\omega_n) + \Gamma_{\qq} +K_{\qq,i\omega_m}
\end{align}
where
\begin{align}
&\vphantom{=}\no\\
\Gamma_{\qq}&=-\delta\Delta_\qq\sigma^1-\delta v_\qq \sigma^3 \no\\
 K_{\qq,i\omega_m}&=-\lambda(\qq,i\omega_m)\sigma^+-\lambda(-\qq,-i\omega_n)^*\sigma^--\xi(\qq,i\omega_m)\sigma^3\\
G^{-1}(\kk,i\omega_n) &= \begin{pmatrix} i\omega_n -\epsilon_\kk+\mu-v_0& -\Delta_0\\- \Delta_0&i\omega_n +\epsilon_\kk-\mu+v_0\end{pmatrix}\no\\\no\\
G(\kk,i\omega_n)&=\left(\begin{array}{cc }%
\frac{u_\kk^2}{i\omega_n-E_\kk}+ \frac{v_\kk^2}{i\omega_n+E_\kk} & u_\kk v_\kk\left[\frac{1}{i\omega_n-E_\kk}-\frac{1}{i\omega_n+E_\kk}\right]\\
                         u_\kk v_\kk\left[\frac{1}{i\omega_n-E_\kk}-\frac{1}{i\omega_n+E_\kk}\right]&\frac{v_\kk^2}{i\omega_n-E_\kk}+ \frac{u_\kk^2}{i\omega_n+E_\kk} \end{array}\right)
\end{align}
where $\omega_n = (2n+1)\pi T$ with integer $n$ is the fermionic Matsubara frequency at temperature $T$. Here $E_\kk=\sqrt{(\epsilon_\kk-\mu+v_0)^2+\Delta_0^2}$ is the Bogoliubov quasiparticle dispersion,
$\epsilon_\kk=-2t (\cos k_x +\cos k_y)$ is the bare band dispersion, and
the BCS coherence factors are given by $u_\kk^2=1-v_\kk^2=(1/2)(1+(\epsilon_\kk-\mu+v_0)/E_\kk)$,
and $u_\kk v_\kk= \Delta_0/2E_\kk$. Here, $\sigma^{i=1,2,3}$ are the Pauli matrices and $\sigma^\pm = \sigma^1\pm i\sigma^2$.
We can now integrate out the fermionic fields to get the action,

\begin{align}
    S=-\text{Tr }\ln\left(\mathcal{G}^{-1}\right)+\frac{1}{U}\left(\Delta_0^2+v_0^2\right)+\frac{1}{U}\sum_{\qq,\omega_m}\left(| \lambda(\qq,i\omega_m)|^2+|\xi_(\qq,i\omega_m)|^2\right)
\end{align}
where we write
\bqa
\text{Tr } \ln\left(\mathcal G^{-1}\right) = \text{Tr } \ln\left( G^{-1}+\Gamma+K\right) = \text{Tr } \ln G^{-1} +\text{Tr } \ln\left(\mathbb I + G\Gamma+G K\right)
\eqa
We then Taylor expand the action about the clean Green's function $G$, with $\Gamma$ and $K$ acting as perturbation. We note that $\Delta_0\delta_{\qq,0}+\delta\Delta_q$ and $v_0\delta_{\qq,0}+\delta v_\qq$ are the exact mean-field static solutions in the presence of disorder. This leads all terms that are linear in $\Gamma_\qq$ to vanish, at all orders of disorder. Moreover, $\delta\Delta_\qq$ and $\delta v_\qq$ are stationary in time and do not transfer any $i\omega_m$. As a result, the coupling between fluctuating fields and static disorder occurs first at the second order in the fluctuating fields and at the first order in disorder. Therefore, the effective action looks like
\bqa
S&=&S_0+S_2+S_3\\
S_0&=& -\text{Tr }\ln\left(G^{-1}\right)+ \frac{N_s\beta}{U}(\Delta_0^2+v_0^2)\\
S_2&=& \frac{1}{2}\sum_{\kk,\qq,\omega_m,\omega_n}\text{Tr }^NG(\kk+\qq,i\omega_n+i\omega_m)K_{\qq,i\omega_m}G(\kk,i\omega_n)K_{-\qq,-i\omega_m} +\frac{1}{U}\sum_{\qq,\omega_m}\left( |\lambda(\qq,i\omega_m)|^2+|\xi_(\qq,i\omega_m)|^2\right)\\
S_3&=&-\frac12\sum_{\kk,\qq,\qq_1,\omega_m,\omega_n}\text{Tr }^NG(\kk+\qq+\qq_1,i\omega_n+i\omega_m)K_{\qq+\qq_1,i\omega_m}G(\kk,i\omega_n)\Gamma_{-\qq_1}G(\kk+\qq_1,i\omega_n)K_{-\qq,-i\omega_m}
\eqa
We note that all summations over momentum indices are normalised by a factor of number of sites $N_s$, while all summations over Matsubara indices are normalised by a factor of inverse temperature $1/T$. Here, $S_2$ represents the Gaussian action of the fluctuations and $S_3$ corresponds to the couplings between the fluctuations and the disorder. $N_s$ is the number of sites in the system. We can simplify $S_2$ to get the clean case fluctuating action as following,

\beq
S_2=\frac{1}{2}\sum_{\qq,\omega_m} \Lambda^\dagger(\qq,i\omega_m) \hat{M}^{-1}(\qq,i\omega_m)\Lambda(\qq,i\omega_m) 
\eeq
where $\Lambda^\dagger(\qq,i\omega_m)=[\lambda^\ast(\qq,i\omega_m),\lambda(-\qq,-i\omega_m),\xi(\qq,i\omega_m)]$ and the components of $\hat{M}^{-1}(\qq,i\omega_m)$ are 
\bqa
\no M^{-1}_{11}(\qq,i\omega_m)&=& \frac{1}{U} + \sum_{\kk,\omega_n} \text{Tr }^NG(\kk+\qq,i\omega_n+i\omega_m)\sigma^+ G(\kk,i\omega_n)\sigma^-=\frac{1}{U}+\sum_\kk(u^2u'^2-v^2v'^2)I(\kk,\qq,i\omega_m)\\
M^{-1}_{12}(\qq,i\omega_m)&=&\sum_{\kk,\omega_n} \text{Tr }^NG(\kk+\qq,i\omega_n+i\omega_m)\sigma^+ G(\kk,i\omega_n)\sigma^+=-\sum_\kk uu'vv' I(\kk,\qq,i\omega_m)\\
\no M^{-1}_{33}(\qq,i\omega_m)&=& \frac{2}{U} +\sum_{\kk,\omega_n} \text{Tr }^NG(\kk+\qq,i\omega_n+i\omega_m)\sigma^3 G(\kk,i\omega_n)\sigma^3=\frac{2}{U}+\sum_{\kk} (uu'+vv')^2I(\kk,\qq,i\omega_m)\\
\no M^{-1}_{13}&=&  \sum_{\kk,\omega_n} \text{Tr }^NG(\kk+\qq,i\omega_n+i\omega_m)\sigma^+ G(\kk,i\omega_n)\sigma^3=-\sum_\kk(u'v+uv')\left( \frac{uu'}{i\omega_m-E-E'}-\frac{vv'}{i\omega_m-E-E'}\right)\\
\no  M^{-1}_{23}&=&  \sum_{\kk,\omega_n} \text{Tr }^NG(\kk+\qq,i\omega_n+i\omega_m)\sigma^- G(\kk,i\omega_n)\sigma^3=\sum_\kk(u'v+uv')\left( \frac{vv'}{i\omega_m-E-E'}-\frac{uu'}{i\omega_m-E-E'}\right)
\eqa
with $M^{-1}_{22}=\left[M^{-1}_{11}\right]^\ast$ and $M^{-1}_{\beta\alpha}=\left[M^{-1}_{\alpha\beta}\right]^\ast$ for $\alpha\neq \beta$. Here, $I(\kk,\qq,i\omega_m)= \frac{1}{i\omega_m-E-E'}-\frac{1}{i\omega_m-E-E'}$, and $u,v,E\equiv u_\kk,v_\kk,E_\kk$ and $u',v',E'\equiv u_{\kk+\qq},v_{\kk+\qq},E_{\kk+\qq}$ respectively. This can be written in a compact notation as 
\bqa
\no M^{-1}_{ij}(\qq,i\omega_m)&=& \frac{\delta_{ij}(1+\delta_{i3})}{U} + \sum_{\kk,\omega_n} \text{Tr }^NG(\kk+\qq,i\omega_n+i\omega_m)\sigma^{\alpha_i} G(\kk,i\omega_n)\sigma^{\beta_j}
\eqa
With $\alpha_i=+,-,3$ for $i=1,2,3$ resepectively and $\beta_j=-,+,3$ for $j=1,2,3$ resepectively. The Matsubara summations can be done using the MatsubaraSum package in Mathematica.

Now we turn to $S_3$ term in the action. $S_3$ can be expanded  to obtain the following form, 
\bqa
S_3&=&\sum_{\qq,\qq_1\omega_m} \Lambda^\dagger(\qq+\qq_1,i\omega_m) \hat{F}(\qq,\qq_1,i\omega_m)\Lambda(\qq,i\omega_m)
\eqa
where, similar to $\hat{M^{-1}}$, we can write the components of $\hat{F}$ as
\bqa
\hat{F}_{ij}(\qq,\qq_1,i\omega_m)&=&f_{ij}(\qq,\qq_1,i\omega_m)\delta \Delta_{-\qq_1}+ g_{ij}(\qq,\qq_1,i\omega_m)\delta v_{-\qq_1}\\
 f_{ij}(\qq,\qq_1,i\omega_m)&=& \sum_{\kk,\omega_n} \text{Tr }^NG(\kk+\qq+\qq_1,i\omega_n+i\omega_m)\sigma^{\alpha_i} G(\kk,i\omega_n)\sigma^{1}G(\kk+\qq_1,i\omega_n)\sigma^{\beta_j}\\
 g_{ij}(\qq,\qq_1,i\omega_m)&=& \sum_{\kk,\omega_n} \text{Tr }^NG(\kk+\qq+\qq_1,i\omega_n+i\omega_m)\sigma^{\alpha_i} G(\kk,i\omega_n)\sigma^{3}G(\kk+\qq_1,i\omega_n)\sigma^{\beta_j}
\eqa
where $\alpha_i$ and $\beta_j$ are defined as before. To evaluate the the trace over the product of $3$ Green's function, we note that each component can be written in the form $\frac{\alpha }{i\omega_n -E}+\frac{\beta }{i\omega_n+E }$ with $E>0$. Then, we use the following formula evaluated at zero temperature,
\begin{align}
   \no&\sum_{i\omega_n} \left(\frac{\alpha }{i\omega_n -E}+\frac{\beta }{i\omega_n+E }\right) \left(\frac{\gamma'' }{i\omega_n -E''}+\frac{\delta'' }{i\omega_n+E'' }\right) \left(\frac{\lambda' }{i\omega_n+iq_n -E'}+\frac{\mu' }{i\omega_n+iq_n +E' }\right)\\
   \no&\hphantom{3}\\
   \no&= -\frac{1}{E+E''}\left( \alpha\delta''\left(\frac{\mu'}{iq_n + E+E'} +\frac{\lambda'}{iq_n -E'-E''}\right)+\beta\gamma''\left(\frac{\mu'}{iq_n + E'+E''} +\frac{\lambda'}{iq_n -E-E'}\right)\right)\\
   &+\frac{1}{E-E''}\left(\alpha\gamma''\mu'\left(\frac{1}{iq_n +E'+E''}-\frac{1}{iq_n +E+E'}\right)+\beta\delta''\lambda'\left(\frac{1}{iq_n-E'-E''}-\frac{1}{iq_n-E-E'}\right)\right)\label{eq:prodmats}
\end{align}
where $\omega_n$ is the fermionic Matsubara frequency and $q_n$ is the bosonic Matsubara frequency. Since the denominators are of the form $iq_n-E-E'$ or $iq_n+E+E'$, after analytical continuation of $iq_n\to \omega+i\eta$, these functions will not have any imaginary part. Therefore, the couplings are purely real.
\section{The 2 Band Model}
\label{app:2band}
In the 2 band model, we assume that the disorder can be modelled by an effective potential of the form $\delta c(\QQ)\sum_r (-1)^rc^\dagger_rc_r$ which represents an effective CDW fluctuation. The Hamiltonian is given by
\begin{align*}
    H&= -t \sum_{\langle rr'\rangle\sigma}c^\dagger_{r\sigma}c_{r'\sigma} - U \sum_r n_{r\uparrow}n_{r\downarrow} +\sum_r (v_r - \mu) n_r
\end{align*}

Proceeding as before, we perform a perturbation series about the clean case superconductor. We get the same clean case Green's function as in  Section~\ref{app:cartesian}. However, for now we assume a fluctuation in the density channel of the form $v^{eff}(r)\to v_0+\delta v(r)+\xi(r,\tau)$ to get

\begin{align}
    \mathcal{G}^{-1}(\kk+\qq,\kk,i\omega_n) &=  \left(i\omega_n \sigma^0 -(\epsilon_\kk-\mu+v_0)\sigma^3-\Delta_0 \sigma^1\right)\delta_{\qq,0} -\delta v_{\qq}\sigma^3- \xi_{\qq}\sigma^3
\end{align}

We now model the fluctuations in the amplitude phase coordinates given by the substitution (We do not assume any static fluctuations in the density channel)
$$\Delta_0\to (\Delta_0+\eta(r,\tau))e^{i\theta(r,\tau)}$$

We can perform a Gauge transformation to eliminate the phase factor to get the transformed Green's function\cite{HiggsAbhisek}. This can be expanded upto second order in $\theta$ to get (Where $\QQ=[\pi,\pi]$)

\begin{align}
\no \tilde{\mathcal{G}}^{-1}(\kk+\qq,\kk,i\omega_m+i\omega_n,i\omega_n) &= G^{-1}(\kk,i\omega_n) +\mathcal{K}(\kk+\qq,\kk,i\omega_m)+V(\kk+\qq,\kk)\\
\no G^{-1}(\kk,i\omega_n)&=  \left(i\omega_n\sigma^0 -\zeta_\kk\sigma^3 -\Delta_0 \sigma^1\right)\delta_{\qq,0} \\
\no \mathcal{K}(\kk+\qq,\kk,i\omega_m)&=\Gamma_1(\kk+\qq,\kk,i\omega_m)+\Gamma_2(\kk+\qq,\kk,i\omega_m)+\Gamma_3(\kk+\qq,\kk,i\omega_m)+K(\kk+\qq,\kk,i\omega_m)\\
\no \Gamma_1(\kk+\qq,\kk,i\omega_m)&= \frac{i}{2}\theta_\qq\left( (i\omega_m\sigma^3 +\left(\zeta_{\kk+\qq}-\zeta_\kk\right)\sigma^0\right)\\
\no \Gamma_2(\kk+\qq,\kk,i\omega_m)&=-\frac{1}{8} \sum_{\qq_1} \theta_{\qq_1}\theta_{\qq-\qq_1} \left(\zeta_\kk+\zeta_{\kk+\qq}-\zeta_{\kk+\qq-\qq_1}-\zeta_{\kk+\qq_1}\right)\sigma^3\\
\no \Gamma_3(\kk+\qq,\kk,i\omega_m)&=- \xi_\qq\sigma^3\\
\no K(\kk+\qq,\kk,i\omega_m)&=-\eta_\qq\sigma^1\\
 V(\kk+\qq_1,\kk)&=-\delta v_{\qq_1}\sigma^3\delta_{m,0}
\end{align}

Where $\epsilon_\kk -\mu+v_0 = \zeta_\kk$. The action is given by

\begin{align}
    S=S_0-\text{Tr }\ln\left(\mathbb I+G\mathcal{K}+GV\right)+ \frac{1}{U}\sum_{\qq,i\omega_m}\left( |\eta_{\qq,i\omega_m}|^2+|\xi_{\qq,i\omega_m}|^2\right)
\end{align}
We can expand the action upto second order in fluctuationg fields and first order in $V$ as in Appendix~\ref{app:cartesian} to get the following.

\begin{equation}
S = S_0 +\frac{1}{U}\sum_{\qq,i\omega_m}\left( |\eta_{\qq,i\omega_m}|^2+|\xi_{\qq,i\omega_m}|^2\right) - \text{Tr }G \mathcal K + \frac12 \text{Tr }G \mathcal K G \mathcal K +\text{Tr }G\Gamma_2GV-\frac{1}{2}\text{Tr }G\mathcal K G\mathcal K GV
\end{equation}

The term linear in $\mathcal K$ contains terms linear in the fluctuations which goes to $0$ because of the mean field equation, while the quadratic term gives us the gaussian couplings. These couplings are given by 

\beq
S_{g}=\sum_{\qq,\omega_m}\Gamma(\qq,i\omega_m)\hat{D}^{-1}(\qq,i\omega_m)\Gamma(-\qq,-i\omega_m)
\eeq
where $\Gamma(\qq,i\omega_m)=[\eta(\qq,i\omega_m),\theta(\qq,i\omega_m),\xi(\qq,i\omega_m)$ is a
three component field containing the amplitude ($\eta$), the phase ($\theta$) and the Hartree potential ($\xi$) fluctuations. The components of $D^{-1}$ are given by 

\bqa
 \no D^{-1}_{11}(\qq,i\omega_m)&=& \frac{1}{U} + \frac12 \sum_{\kk,i\omega_n}\text{Tr }^N G(\kk+\qq,i\omega_n+i\omega_m)\sigma^1G(\kk,i\omega_n)\sigma^1\\
\no &=& \frac{1}{U}+ \frac12 \sum_\kk (uu'-vv')^2I(\kk,\qq,i\omega_m)\\
\no D^{-1}_{12}(\qq,i\omega_m)&=& \frac{i}{4}(i\omega_m)\sum_{\kk,i\omega_n}\text{Tr }^N  G(\kk+\qq,i\omega_n+i\omega_m)\sigma^1G(\kk,i\omega_n)\sigma^3 = -D^{-1}_{12}(\qq,i\omega_m)\\
\no &=&- \frac{i}{4}(i\omega_m)\sum_\kk (uv'+vu')(uu'-vv')I(\kk,\qq,i\omega_m)\\
 \no D^{-1}_{22}(\qq,i\omega_m)&=& (i\omega_m)^2 \kappa(\qq,i\omega_m)+\sum_{\hat\delta=\pm \hat x,\hat y}(1-\cos(\qq\cdot \hat\delta)\Theta_\delta+\chi(\qq,i\omega_m)\\
\no D^{-1}_{13}(\qq,i\omega_m) &=& \frac12 \sum_{\kk,i\omega_n}\text{Tr }^N G(\kk+\qq,i\omega_n+i\omega_m)\sigma^1G(\kk,i\omega_n)\sigma^3 = D^{-1}_{31}(\qq,i\omega_m)\\
\no &=& -\frac12 \sum_\kk (uv'+vu')(uu'-vv')I(\kk,\qq,i\omega_m)\\
 \no D^{-1}_{23}(\qq,i\omega_m)&=&-\frac{i}{4}(i\omega_m)\sum_{\kk,i\omega_n}\text{Tr }^N G(\kk+\qq,i\omega_n+i\omega_m)\sigma^3G(\kk,i\omega_n)\sigma^3=-D^{-1}_{32}(\qq,i\omega_m)\\
\no &=&-\frac{i}{4}(i\omega_m) \sum_\kk (uv'+vu')^2I(\kk,\qq,i\omega_m)\\
\no D^{-1}_{33}(\qq,i\omega_m)& =& \frac{1}{U}+\frac12 \sum_{\kk,i\omega_n}\text{Tr }^N G(\kk+\qq,i\omega_n+i\omega_m)\sigma^3G(\kk,i\omega_n)\sigma^3\\
&=&\frac{1}{U}+ \frac12\sum_\kk (uv'+vu')^2I(\kk,\qq,i\omega_m)
\eqa
where $\kappa$ is the generalized compressibility and $\Theta_\delta$ is the current-current correlator in the system given by
\bqa
\no \kappa(\qq,i\omega_m) &=& \frac18 \sum_{\kk,i\omega_n}\text{Tr }^N G(\kk+\qq,i\omega_n+i\omega_m)\sigma^3G(\kk,i\omega_n)\sigma^3\\
\no &=&\frac18 \sum_\kk (uv'+vu')^2I(\kk,\qq,i\omega_m)\\
\no \Theta_\delta &=& \frac{t}{2}\sum_{E_\kk >0}v_\kk^2\cos(\kk\cdot \hat\delta)\\
 \no \chi(\qq,i\omega_m) &=& \frac{t^2}{8}\sum_\kk (\zeta_{\kk+\qq} - \zeta_\kk)^2 \sum_{i\omega_n}\text{Tr }^N G(\kk+\qq,i\omega_n+i\omega_m)\sigma^0G(\kk,i\omega_n)\sigma^0\\
\no &=& \frac{t^2}{8}\sum_\kk (\zeta_{\kk+\qq}-\zeta_\kk)^2(uv'-vu')^2I(\kk,\qq,i\omega_m)\\
I(\kk,\qq,i\omega_m) &=& \frac{1}{i\omega_m-E-E'}-\frac{1}{i\omega_m+E+E'}
\eqa

Here, $u,v,E\equiv u_\kk,v_\kk,E_\kk$ and $u',v',E'\equiv u_{\kk+\qq},v_{\kk+\qq},E_{\kk+\qq}$ respectively. Next, we investigate only the disorder couplings. We will define $\tilde{V} =G(\kk+\qq+\qq_1,i\omega_m+i\omega_n) V(\kk+\qq+\qq_1,\kk+\qq)$. Next, we analyse the terms of the action that correspond to the scattering of $\eta,\theta,\xi$ fields by the disorder potential $V$. 

There are two terms which goes as $\sim \theta \theta\delta v$. The term coming from $\text{Tr }G\Gamma_2GV$ is given by
\begin{align}
    &\sum_{\kk,\qq_1,\qq,\omega_n,\omega_m}\text{Tr }^N G(\kk,i\omega_n)\Gamma_2(\kk,\kk+\qq_1) G(\kk+\qq_1) V(\kk+\qq_1,\kk) \\
    &\no=-\sum_{\qq,\qq_1,i\omega_m}\theta_{\qq}\theta_{-\qq-\qq_1}\delta v_{\qq_1}\sum_{\kk,i\omega_n}\left(-\frac{1}{8}  \left(\zeta_{\kk+\qq_1}+\zeta_{\kk}-\zeta_{\kk-\qq}-\zeta_{\kk+\qq_1+\qq}\right)\right)\text{Tr }^NG(\kk,i\omega_n)\sigma^3G(\kk+\qq_1)\sigma^3
\end{align}
The term coming from $\text{Tr }G\Gamma_1G\Gamma_1GV$ is given by
\begin{align}
    &\sum_{\kk,\qq_1,\qq,i\omega_n,i\omega_m}\text{Tr }^N G(\kk+\qq,i\omega_m+i\omega_n) \Gamma_1(\kk+\qq,\kk,i\omega_m) G(\kk,i\omega_n) \Gamma_1(\kk,\kk+\qq+\qq_1,-i\omega_m)\tilde{V}\\
    &\no=-\sum_{\qq_1,\qq,i\omega_m} \frac{-1}{4}\theta_\qq\theta_{-\qq-\qq_1}\delta v_{\qq_1}\sum_{\kk,i\omega_n}(-(i\omega_m)^2\text{Tr }^N G'\sigma^3G\sigma^3G''\sigma^3 +(i\omega_m)(\zeta-\zeta'')\text{Tr }^NG_0'\sigma^3G_0G_0''\sigma^3\\
    &\no-(i\omega_m)(\zeta'-\zeta)\text{Tr }^NG'G\sigma^3G''\sigma^3+ (\zeta'-\zeta)(\zeta-\zeta'')\text{Tr }^NG'GG''\sigma^3)\end{align}
Where $G\equiv G(\kk,i\omega_n), G'\equiv G(\kk+\qq,i\omega_n+i\omega_m), G''\equiv G(\kk+\qq+\qq_1,i\omega_m+i\omega_n)$ and $\zeta\equiv \zeta_{\kk}, \zeta'\equiv \zeta_{\kk+\qq},\zeta''\equiv \zeta_{\kk+\qq+\qq_1}$. The term coming from $\text{Tr }G\Gamma_1GKGV$ is given by
\begin{align}
    &\sum_{\kk,\qq_1,\qq,i\omega_n,i\omega_m}\text{Tr }^N G(\kk+\qq,i\omega_m+i\omega_n) \Gamma_1(\kk+\qq,\kk,i\omega_m) G(\kk,i\omega_n) K(\kk,\kk+\qq+\qq_1,-i\omega_m)\tilde{V}\\
    &\no =-\sum_{\qq_1,\qq,i\omega_n}\frac{-i}{2}\theta_\qq\eta_{-\qq-\qq_1}\delta v_{\qq_1}\sum_{\kk,i\omega_n} \left((i\omega_m)\text{Tr }^NG'\sigma^3G\sigma^1G''\sigma^3+ (\zeta'-\zeta)\text{Tr }^NG'G\sigma^1G''\sigma^3\right)
   \end{align}
The term coming from $\text{Tr }GKG\Gamma_1GV$ is given by
\begin{align}
    &\sum_{\kk,\qq_1,\qq,i\omega_m,i\omega_n}\text{Tr }^N G(\kk+\qq,i\omega_m+i\omega_n) K(\kk+\qq,\kk,i\omega_m) G(\kk,i\omega_n) \Gamma_1(\kk,\kk+\qq+\qq_1,-i\omega_m) \tilde{V}\\
    &\no =-\sum_{\qq_1,\qq,i\omega_m}\frac{-i}{2}\eta_\qq\theta_{-\qq-\qq_1}\delta v_{\qq_1}\sum_{\kk,i\omega_n} \left((-i\omega_m)\text{Tr }^NG'\sigma^1G\sigma^3G''\sigma^3+ (\zeta-\zeta'')\text{Tr }^NG'\sigma^1GG''\sigma^3\right)
  \end{align}
The term coming from $\text{Tr }GKGKGV$ is given by
\begin{align}
    &\sum_{\kk,\qq_1,\qq,i\qq}\text{Tr }^N G(\kk+\qq,i\omega_m+i\omega_n) K(\kk+\qq,\kk,i\omega_m) G(\kk,i\omega_n) K(\kk,\kk+\qq+\qq_1,-i\omega_m)\tilde{V}\\
    &\no =-\sum_{\qq_1,\qq,i\omega_m}\eta_\qq\eta_{-\qq-\qq_1}\delta v_{\qq_1}\sum_{\kk,i\omega_n} \text{Tr }^NG'\sigma^1G\sigma^1G''\sigma^3
\end{align}
The terms coming from $\text{Tr } G\Gamma_1 G\Gamma_3GV$ is given by
\begin{align}
    &\sum_{\kk,\qq_1,\qq,i\omega_m,i\omega_n}\text{Tr }^N G(\kk+\qq,i\omega_m+i\omega_n) \Gamma_1(\kk+\qq,\kk,i\omega_m) G(\kk,i\omega_n) \Gamma_3(\kk,\kk+\qq+\qq_1,-i\omega_m)\tilde{V}\\
    &\no =-\sum_{\qq_1,\qq,i\omega_m} \frac{-i}{2}\theta_\qq\xi_{-\qq-\qq_1}\delta v_{\qq_1}\sum_{\kk,i\omega_n}((i\omega_m)\text{Tr }^N G'\sigma^3G\sigma^3G''\sigma^3 +(\zeta'-\zeta)\text{Tr }^NG'G\sigma^3G''\sigma^3)
    \end{align}
The term coming from $\text{Tr }G\Gamma_3G\Gamma_1GV$ is given by
\begin{align}
    &\sum_{\kk,\qq_1,\qq,i\omega_m,i\omega_n}\text{Tr }^N G(\kk+\qq,i\omega_m+i\omega_n) \Gamma_3(\kk+\qq,\kk,i\omega_m) G(\kk,i\omega_n) \Gamma_1(\kk,\kk+\qq+\qq_1,-i\omega_m)\tilde{V}\\
    &\no =-\sum_{\qq_1,\qq,i\omega_m} \frac{-i}{2}\xi_\qq\theta_{-\qq-\qq_1}\delta v_{\qq_1}\sum_{\kk,i\omega_n}((-i\omega_m)\text{Tr }^N G'\sigma^3G\sigma^3G''\sigma^3 +(\zeta-\zeta'')\text{Tr }^NG'\sigma^3GG''\sigma^3)
\end{align}
The term coming from $\text{Tr }G\Gamma_3GKGV$ is given by
\begin{align}
    &\sum_{\kk,\qq_1,\qq,i\omega_m,i\omega_n}\text{Tr }^N G(\kk+\qq,i\omega_m+i\omega_n) \Gamma_3(\kk+\qq,\kk,i\omega_m) G(\kk,i\omega_n) K(\kk,\kk+\qq+\qq_1,-i\omega_m)\tilde{V}\\
    &\no =-\sum_{\qq_1,\qq,i\omega_m}\xi_\qq\eta_{-\qq-\qq_1}\delta v_{\qq_1}\sum_{\kk,i\omega_n} \text{Tr }^NG'\sigma^3G\sigma^1G''\sigma^3
\end{align}
The term coming from $\text{Tr }GKG\Gamma_3GV$ is given by
\begin{align}
    &\sum_{\kk,\qq_1,\qq,i\omega_m,i\omega_n}\text{Tr }^N G(\kk+\qq,i\omega_m+i\omega_n)K(\kk+\qq,\kk,i\omega_m) G(\kk,i\omega_n)  \Gamma_3(\kk,\kk+\qq+\qq_1,-i\omega_m) \tilde{V}\\
    &\no =-\sum_{\qq_1,\qq,i\omega_m}\eta_\qq\xi_{-\qq-\qq_1}\delta v_{\qq_1}\sum_{\kk,i\omega_n} \text{Tr }^NG'\sigma^1G\sigma^3G''\sigma^3
\end{align}
The term coming from $\text{Tr }G\Gamma_3G\Gamma_3GV$ is given by
\begin{align}
     &\sum_{\kk,\qq_1,\qq,i\omega_m,i\omega_n}\text{Tr }^N G(\kk+\qq,i\omega_m+i\omega_n) \Gamma_3(\kk+\qq,\kk,i\omega_m) G(\kk,i\omega_n) \Gamma_3(\kk,\kk+\qq+\qq_1,-i\omega_m)\tilde{V}\\
    &\no =-\sum_{\qq_1,\qq,i\omega_m} \xi_\qq\xi_{-\qq-\qq_1}\delta v_{\qq_1}\sum_{\kk,i\omega_n}\text{Tr }^NG'\sigma^3 G\sigma^3G''\sigma^3
\end{align}

Considering only the case of $\delta v_{q_1}=\delta v(\QQ)\delta_{\qq_1,\QQ}$, the disorder coupling can be written as 
\beq
S_{fl}=\frac{1}{2}\sum_{\qq,\omega_m} [\Gamma(\qq,i\omega_m),\Gamma(\qq-\QQ,i\omega_m)] \left[ \begin{array}{cc}
                                                                                                   \hat{D}^{-1}(\qq,i\omega_m) & \delta v(\QQ) \hat{F}(\qq,i\omega_m)\\
                                                                                                   \delta v(\QQ)^\ast \hat{F}(\qq+\QQ,i \omega_m) &\hat{D}^{-1}(\qq+\QQ,i\omega_m)
                                                                                                 \end{array}\right] \left[ \begin{array}{c} \Gamma(-\qq,-i\omega_m)\\
                                                                                                                   \Gamma(-\qq+\QQ,-i\omega_m)        \end{array}\right]
\eeq
where $\Gamma(\qq,i\omega_m)=[\eta(\qq,i\omega_m),\theta(\qq,i\omega_m),\xi(\qq,i\omega_m)]$ is a
three component field containing the amplitude ($\eta$), the phase ($\theta$) and the Hartree potential ($\xi$) fluctuations. $F(\qq,i\omega_m)$ is a $3\times 3$ matrix that couples the modes at $\qq$ with the mode at $-\qq+\QQ$, and is first order in $M$. We can write the off diagonal coupling term as

\begin{equation}
\delta v(\QQ)\sum_{\qq\qq,i\omega_m} \left(\begin{array}{cccc}\eta_{\qq}&
\theta_\qq&\xi_\qq
\end{array}\right)
\left(\begin{array}{ccc}
F_{11} &F_{12}&F_{13}\\
F_{21} &F_{22}&F_{23}\\
F_{31} &F_{32}&F_{33}\\
\end{array}\right)
\left(\begin{array}{cc}\eta_{-\qq+\QQ}\\
\theta_{-\qq+\QQ}\\\xi_{-\qq+\QQ}
\end{array}\right),
\end{equation}
Where the coefficients are given by 
\begin{eqnarray}
    \no F_{11}&=&\frac12 \Pi_{113}\\
    \no F_{21}&=&-\frac{i}{4} \left( (i\omega_m) \Pi_{313} +\left[ \zeta_{\kk+\qq}-\zeta_\kk\right]\Pi_{013}\right)\\
    \no F_{12}&=&-\frac{i}{4} \left( (-i\omega_m) \Pi_{133} +\left[ \zeta_{\kk}-\zeta_{\kk+\qq+\QQ} \right]\Pi_{013}\right)\\
    \no F_{22}&=& \frac{(i\omega_m)^2}{8}\Pi_{333}-\frac{i\omega_m}{8}\left[\zeta_\kk-\zeta_{\kk+\qq+\QQ}\right]\Pi_{303}+\frac{i\omega_m}{8}\left[\zeta_{\kk+\qq}-\zeta_{\kk}\right]\Pi_{033} \\
   \no  &&+f_{22}^0 -\left[ \frac{(\zeta_{\kk+\qq}-\zeta_\kk)(\zeta_\kk-\zeta_{\kk+\qq+\QQ})}{8}\right]\Pi_{003} \\
    \no F_{13}&=&\frac12\Pi_{133}\\
    \no F_{23}&=&-\frac{i}{4} (i\omega_m)\Pi_{333}-\frac{i}{4}\left[\zeta_{\kk+\qq}-\zeta_\kk\right]\Pi_{033}\\
   \no  F_{31}&=&-\frac{1}{2}\Pi_{313}\\
    \no F_{32}&=&-\frac{i}{4}(-i\omega_m)\Pi_{333} - \frac{i}{4}\left[\zeta_\kk-\zeta_{\kk+\qq+\QQ}\right]\Pi_{303}\\
     F_{33}&=&\frac{1}{2}\Pi_{333}
\end{eqnarray}
Where we've defined 

\begin{eqnarray}
   \no \left[ f(\kk,\qq)\right] \Pi_{abc}(q,iq_n) &=& \sum_{\kk}f(\kk,\qq)\sum_{i\omega_n} \text{Tr }^NG(\kk+\qq,i\omega_n+i\omega_m) \sigma^a G(\kk,i\omega_n)\sigma^b G(\kk+\qq+\QQ,i\omega_n+i\omega_m) \sigma^c\\
   \no  f_{22}^0&=&\sum_{\kk,i\omega_n} (\zeta_{\kk+\qq}-\zeta_{\kk-\qq}) \text{Tr }^N G(\kk,i\omega_n)\sigma^zG(\kk+\QQ,i\omega_n)\sigma^z\\
&=& \sum_\kk (\zeta_{\kk+\qq}-\zeta_{\kk-\qq})(uv'+vu')^2I(\kk,\qq,i\omega_m)
\end{eqnarray}
Where $\sigma^a$ are the Pauli matrices for $a=1,2,3$ and $\sigma^0$ is the $2\times 2$ identity matrix. Note that terms that  contain linear factors of the form $\zeta_{k_1}-\zeta_{k_2}$ must be $0$, because they induce a finite current in our system. To evaluate the Matsubara sum of product of 3 Green's functions, we use Eq~\ref{eq:prodmats}.
\end{widetext}

\bibliographystyle{unsrt}
\bibliography{refs}
\end{document}